\documentstyle[psfig,epsfig]{mn}

\newif\ifAMStwofonts

\ifoldfss
  \ifCUPmtlplainloaded \else
    \NewTextAlphabet{textbfit} {cmbxti10} {}
    \NewTextAlphabet{textbfss} {cmssbx10} {}
    \NewMathAlphabet{mathbfit} {cmbxti10} {} 
    \NewMathAlphabet{mathbfss} {cmssbx10} {} 
  \fi
  \ifAMStwofonts
    \ifCUPmtlplainloaded \else
      \NewSymbolFont{upmath} {eurm10}
      \NewSymbolFont{AMSa} {msam10}
      \NewMathSymbol{\upi}     {0}{upmath}{19}
      \NewMathSymbol{\umu}     {0}{upmath}{16}-----------------------------------------------------------------------

      \NewMathSymbol{\upartial}{0}{upmath}{40}
      \NewMathSymbol{\leqslant}{3}{AMSa}{36}
      \NewMathSymbol{\geqslant}{3}{AMSa}{3E}

       \let\le=\leqslant
       \let\ge=\geqslant
    \fi
  \fi
\fi 

\ifnfssone
  \newmathalphabet{\mathit}
  \addtoversion{normal}{\mathit}{cmr}{m}{it}
  \addtoversion{bold}{\mathit}{cmr}{bx}{it}
  \newmathalphabet{\mathbfit} 
  \addtoversion{normal}{\mathbfit}{cmr}{bx}{it}
  \addtoversion{bold}{\mathbfit}{cmr}{bx}{it}
  \newmathalphabet{\mathbfss} 
  \addtoversion{normal}{\mathbfss}{cmss}{bx}{n}
  \addtoversion{bold}{\mathbfss}{cmss}{bx}{n}
  \ifAMStwofonts
    \ifCUPmtlplainloaded \else
      \UseAMStwoboldmath
      \makeatletter
      \new@mathgroup\upmath@group
      \define@mathgroup\mv@normal\upmath@group{eur}{m}{n}
      \define@mathgroup\mv@bold\upmath@group{eur}{b}{n}
      \edef\UPM{\hexnumber\upmath@group}
      \new@mathgroup\amsa@group
      \define@mathgroup\mv@normal\amsa@group{msa}{m}{n}
      \define@mathgroup\mv@bold\amsa@group{msa}{m}{n}
      \edef\AMSa{\hexnumber\amsa@group}
      \makeatother
      \mathchardef\upi="0\UPM19
      \mathchardef\umu="0\UPM16
      \mathchardef\upartial="0\UPM40
      \mathchardef\leqslant="3\AMSa36
      \mathchardef\geqslant="3\AMSa3E

       \let\le=\leqslant
       \let\ge=\geqslant
    \fi
  \fi
\fi 

\ifnfsstwo
  \DeclareMathAlphabet{\mathbfit}{OT1}{cmr}{bx}{it}
  \SetMathAlphabet\mathbfit{bold}{OT1}{cmr}{bx}{it}
  \DeclareMathAlphabet{\mathbfss}{OT1}{cmss}{bx}{n}
  \SetMathAlphabet\mathbfss{bold}{OT1}{cmss}{bx}{n}
  \ifAMStwofonts
    \ifCUPmtlplainloaded \else
      \DeclareSymbolFont{UPM}{U}{eur}{m}{n}
      \SetSymbolFont{UPM}{bold}{U}{eur}{b}{n}
      \DeclareSymbolFont{AMSa}{U}{msa}{m}{n}
      \DeclareMathSymbol{\upi}{0}{UPM}{"19}
      \DeclareMathSymbol{\umu}{0}{UPM}{"16}
      \DeclareMathSymbol{\upartial}{0}{UPM}{"40}
      \DeclareMathSymbol{\leqslant}{3}{AMSa}{"36}
      \DeclareMathSymbol{\geqslant}{3}{AMSa}{"3E}

       \let\le=\leqslant
       \let\ge=\geqslant
    \fi
  \fi
\fi 

\ifCUPmtlplainloaded \else
  \ifAMStwofonts \else 
    \def\upi{\pi}
    \def\umu{\mu}
    \def\upartial{\partial}
  \fi
\fi
\newcommand{\beq}{\begin{equation}}
\newcommand{\eeq}{\end{equation}}
\newcommand{\bey}{\begin{eqnarray}}
\newcommand{\eey}{\end{eqnarray}}
\newcommand{\pc}{\, {\rm pc} }
\newcommand{\kpc}{\, {\rm kpc} }

\newcommand{\msun}{M_\odot}

\newcommand{\lsun}{L_\odot}

\newcommand{\grad}{{\bf \nabla}}

\newcommand{\be}{\begin{equation}}
\newcommand{\ee}{\end{equation}}
\newcommand{\ba}{\begin{eqnarray}}
\newcommand{\ea}{\end{eqnarray}}
\newcommand{\nn}{\nonumber \\}

\newcommand{\kms}{\, {\rm km}s^{-1}}
\newcommand{\r}{\mbox{\boldmath $r$}}

\newcommand{\thetab}{\mbox{\boldmath $\theta$}}

\newcommand{\de}{\partial}

\newcommand{\nablab}{\mbox{\boldmath $\nabla$}}

\newcommand{\x}{\mbox{\boldmath $x$}}
\newcommand{\y}{\mbox{\boldmath $y$}}

\newcommand{\barkappa}{\overline{\kappa}}
\newcommand{\baralpha}{\overline{\alpha}}
\newcommand{\tg}{\tilde{g}}
\newcommand{\g}{\mbox{\boldmath $g$}}
\newcommand{\R}{\mbox{\boldmath $R$}}

\title[Testing MOND with Lensing Data] {
 Testing Bekenstein's Relativistic MOND with Lensing Data }
\author[H. Zhao, D.J. Bacon, A.N. Taylor, K. Horne]{
    HongSheng Zhao$^1$, David J. Bacon$^2$, Andy N. Taylor$^2$, Keith Horne$^1$  \\
 $^1$ SUPA, University of St. Andrews, North Haugh, Fife, KY16 9SS, UK \\
 $^2$ SUPA, University of Edinburgh, Royal Observatory, Edinburgh, EH9 3HJ, UK\\
 $^*$email: hz4@st-andrews.ac.uk, djb@roe.ac.uk, ant@roe.ac.uk, kdh1@st-andrews.ac.uk} \date{\today}

\pagerange{\pageref{firstpage}--\pageref{lastpage}}
\pubyear{2006}

\begin{document}

\maketitle

\label{firstpage}

\begin{abstract}
We propose to use multiple-imaged gravitational lenses to set limits on 
gravity theories without dark matter, specificly TeVeS (Bekenstein 2004),  
a theory which is consistent with fundamental relativistic principles 
and the phenomenology of MOdified Newtonian Dynamics (MOND) theory.
After setting the framework for lensing and cosmology, 
we derive analytically the deflection angle for the point lens and the Hernquist
galaxy profile, and study their patterns in convergence, shear and
amplification.  Applying our analytical lensing models we fit galaxy-quasar
lenses in the CASTLES sample.  We do this with three methods, fitting
the observed Einstein ring sizes, the image positions, or the flux
ratios.  In all cases we consistently find that stars in galaxies in  
MOND/TeVeS provide adequate lensing.  Bekenstein's toy
$\mu$ function provides more efficient lensing than 
the standard MOND $\mu$ function.
But for a handful of lenses a good fit would require
a lens mass orders of magnitude larger/smaller than the stellar mass
derived from luminosity unless the modification function $\mu$ and
modification scale $a_0$ for the universal gravity were allowed to be
very different from what spiral galaxy rotation curves normally imply.
We discuss the limitation of present data and summarize 
constraints on the MOND $\mu$ function.  
We also show that the simplest TeVeS "minimal-matter" cosmology,
a baryonic universe with a cosmological constant,  
can fit the distance-redshift relation from the supernova data,
but underpredicts the sound horizon size at the last scattering.  
We conclude that lensing is
a promising approach to differentiate laws of gravity 
(see also astro-ph/0512425).  
\end{abstract}


\section{Introduction}
\par

The standard paradigm of Einsteinian gravity with dark matter and
dark energy has proven amazingly successful at describing the
Universe, in particular the data from the Cosmic Microwave
Background (e.g. Spergel et al 2003), galaxy redshift surveys
(Percival et al 2002, Tegmark et al 2004), Type Ia supernovae
(Reiss et al 1998, Perlmutter et al 1999), and weak gravitational
lensing (see e.g. van Waerbeke \& Mellier 2003, Refregier 2003 for
reviews) to high accuracy with a small number of parameters.
However, it is worth exploring alternative models of gravity to
assess the uniqueness of the model and to open up new ways to
unify gravity and the standard model. Indeed, the detection of
deviations from the Einstein-Hilbert action could signal new
physics, and such deviations are expected, for example, in models
such as the M-theory inspired braneworld.

The central role of both dark matter and dark energy in the
cosmological model has also led some workers to question the
standard paradigm. Given this it seems sensible to develop methods
to test the basic assumptions of this paradigm, if only to put
them on a firmer basis. One of the most direct ways to probe
gravity over large scales in the Universe is via the effect of
gravitational lensing.

In this paper we shall explore gravitational lensing in the
recently developed relativistic version of Modified Newtonian
Dynamics (MOND), the Tensor-Vector-Scalar (TeVeS) theory,
developed by Bekenstein (2004). The non-relativistic version of
MOND was originally proposed by Milgrom (1983) as an alternative
to the dark matter paradigm. Milgrom (1983) suggested that galaxy
rotation curves $V(r)$ could be explained by modifying gravity:
 \be
 g=\frac{V^2}{r}= \frac{GM}{r^2\mu},
~\mu =\cases{ 
{g \over a_0}  & $g \ll a_0 \approx 1.2 \times 10^{-8} {\rm cm}s^{-2}$.\cr
1              &  $g \gg a_0$, \cr}
        \label{mondstep}\label{a0}
 \ee
where the interpolation function 
$\mu(g/a_0)$ is the effective ``dielectric constant", which
itself has the above asymptotic dependence on the gravitational field
strength, $g$.
Thus the gravitational field strength $g$ becomes significantly
stronger than Newtonian gravity $GM/r^2$ in the weak MOND regime,
when $g \le a_0 $.  
 
From a theoretical point of view, MOND has not just one extra free
parameter, $a_0$, which is tuned to explain Tully-Fisher relations, but also
a free interpolation function, $\mu(g/a_0)$.  With a standard choice 
$\mu(x)={x \over \sqrt{1+x^2}}$, one can achieve very good fits to 
contemporary kinematic data of a wide
variety of high and low surface brightness spiral and elliptical
galaxies; even the fine details of velocity curves are reproduced
without fine tuning of the baryonic model (Sanders \& McGaugh,
2002; Milgrom \& Sanders, 2003). However, whether MOND qualifies
as a gravitational theory depends on its prediction of 
fundamental properties of gravity, e.g., the bending of the light.

In a more rigorous treatment of MOND (Bekenstein \& Milgrom,
1984), gravity is the gradient of a conserved potential,
$\Phi(\r)$.  It satisfies a modified Poisson's equation, 
and trajectories of massive particles are governed by the
equation of motion as follows,
 \be \label{binary}
    {d^2 \r \over dt^2} = \g= - \nablab \Phi(\r), 
 \qquad \nablab . \left[\g \mu(g/a_0)\right] = -4\pi G \rho,
 \ee
where the right-hand equation is the density of all baryonic
particles, $g=|\g|$ is the magnitude of the gravitational field.

These versions of MOND, however, all suffer from being
non-relativistic. The main problem is that the theory is not
generally covariant, as the physics still depends on the measured local
acceleration. In the absence of a relativistic version of MOND,
the paradigm could not be used to build cosmological models and
could not provide robust predictions for the expanding Universe,
the Cosmic Microwave Background, the evolution of perturbations,
and gravitational lensing.

Recently a fully relativistic, generally covariant version of MOND
has been proposed, TeVeS (Bekenstein 2004), which passes standard
local and cosmological tests used to check General Relativity. In
this relativistic version, the conformal freedom of a general
relativistic model is used, along with a new scalar field, one
vector field and a conformally coupled metric tensor, to preserve
general covariance. Following Bekenstein's (2004) paper, a number
of other works have appeared studying the cosmological model (Hao
\& Akhoury, 2005) and large-scale structure of the Universe
(Skordis et al., 2005) in the relativistic TeVeS theory.
There are also attempts to refine the Lagrangian for the scalar field 
in TeVeS (hence the MOND interpolation $\mu$-function) 
to improve fits to galaxy rotation curves (Famaey \& Binney 2005) 
and avoid unphysical dilation effects in two-body systems (Zhao \& Famaey 2005).  
A further extention of TeVeS, dubbed BSTV, 
has also been recently developed by Sanders (2005), which 
adds more flexibilities into the theory by making one of the 
non-dynamical scalar field in TeVeS a dynamical field.  It remains to be
seen what the range of predictions of these theories are in cosmology
and solar system dynamics.  

Our goal here is to develop and test TeVeS by making concrete
predictions of lensing in the theory, and comparing these
predictions to data from strong lensing.  We present the general framework
for lensing in TeVeS.  As a first application to data, 
we apply models of simple axisymmetric lenses: 
the point-mass lens and the Hernquist profile. 
While the point-mass is a very poor model for galaxies in
a dark matter theory, it is a reasonable model for lensing in
TeVeS, where the baryons are concentrated in a central region, and
the Einstein ring in most strong lenses encloses most of the
baryons of the galaxy. As an extension to this, we also consider
the Hernquist profile, which allows us to develop our analysis to
include extended galaxies.  

Our lensing formulation bears similarities with that of Qin et al.
(1995), who developed a heuristic formulation to compute lensing
in MOND in the weak field regime. Here we develop a more rigorous
approach based on the relativistic TeVeS. In the weak-field regime
of Bekenstein's model we find a bending angle a factor of two
greater than that found by Qin et al. (1995), consistent with GR.

As reviewed by Bekenstein (2004) and Sanders (2005), there are probably many ways of
generalising MOND to a relativistic theory, with TeVeS (or BSTV) being the
most successful one so far.  Although we derive our results within
the framework of the TeVeS theory, we explore a regime where dynamics behaves 
like in MOND.  Therefore our results are likely relavent for other 
relativistic versions of MOND as well.  

Nevertheless fundamental differences exist between MOND and TeVeS, 
not only at the conceptual level, but also 
at the phenomenology level, particularly when dealing with a non-spherical potential
with tide or external field.  
For example, the Roche lobe of a satellite in MOND is more squashed 
in MOND than naive extrapolations of the Newtonian result (Zhao 2005), and the shape 
depends on the MOND function $\mu$ (Zhao \& Tian 2005) or the TeVeS $\mu$ for the scalar 
field (Zhao \& Famaey 2005).  Major development is also made in 
numerical tools to solve MONDian Poisson's equations 
for studying systems of a general geometry in detail (Ciotti et al. 2005).  
For these reasons, we prefer to follow the TeVeS formulation 
of lensing and cosmology instead of previous 
phenomenology-based formulations (Qin et al. 1995,  Mortlock \& Turner 2001).

The paper is set out as follows. After a brief presentation of the
basic equations of Bekenstein's TeVeS theory in \S2.  We give in
\S3 the central equations for gravitational lensing, and show how
to calculate gravitational potential around a galaxy. An
eager reader could go directly to \S4, where we 
model distances and ages in a homogeneous and isotropic, expanding universe, and
determine the parameters for the minimal-matter cosmology by fitting the high-z SNe
distances. In \S5 we derive the effects of lensing by a point-mass
lens. This model is then generalised to a Hernquist profile in \S6
and applied to observed galaxy-quasar lenses in \S7. We present
our conclusions in \S8.

\section{The TeVeS Equations And Approximations For Galaxies}

Bekenstein's theory involves two metrics, one of which we denote
by $g_{\mu\nu}$ and is the metric in the Einstein frame, and the
other is the physical metric which couples to matter,
$\tg_{\mu\nu}$, following the notation of Bekenstein. The two
metrics (and their inverses) are related by
 \be
    \tg_{\mu\nu} = e^{-2 \phi} g_{\mu\nu} + (e^{-2\phi}-e^{2\phi})U_\mu U_\nu,
 \ee
where $\phi$ is a dynamical scalar field and $U^\mu \equiv
g^{\mu\nu}U_{\nu}$ is a dynamical, time-like vector field,
normalised by $ g^{\mu\nu}U_{\nu}U_{\mu}=1$.
Note that $\tg_{\mu\nu} = g_{\mu\nu}$ in the limit that the scalar
field $\phi=0$.  The dynamics of the vector field are governed by
an action (cf. eq. 26 of Bekenstein), but for the problems involving
lensing, which deal with either a static galaxy or a homogeneous
universe, the vector field simplifies in the Einstein frame to
 \beq
        U^\mu = [(-g_{tt})^{-{1 \over 2}}c,0,0,0],
 \eeq
i.e., a four-vector which is parallel to the time axis apart but
with a different normalisation. The gravity sector of the theory
is given by the Einstein-Hilbert action in the Einstein frame,
  \be
    S_g = \int \! d^4x \, \sqrt{-g}
    g^{\mu\nu} \left( { R_{\mu\nu} \over 16 \pi G_K} +
    {\rho_\Lambda c^2 g_{\mu\nu} \over 2} \right),
  \ee
where $G_K$ is a parameter of the theory, and  here $g$ is the
determinant of the metric tensor, $R_{\mu\nu}$ is the Ricci tensor
in the Einstein frame, and $\rho_\Lambda c^2$ is the energy
density due to a cosmological constant. Matter is coupled to the
physical metric by the usual action $S_m = \int \! d^4 x\, \sqrt{-\tg}{\cal L}_m$
where ${\cal L}_m$ is the Lagrangian for the luminous matter
fields.  Hence the physical energy momentum tensor
 \beq
   \tilde{T}_{\mu\nu} =
    \left[(\tilde{\rho} c^2+\tilde{p}) \tilde{u}_\mu \tilde{u}_\nu + \tilde{p} \tg_{\mu\nu} \right],
    \qquad \tilde{u}_\mu={\rm e}^{\phi}U_\mu,
   \label{enmomtensor}
 \eeq
where the renormalised physical four-velocity, $\tilde{u}_\mu$, is
treated as colinear with the time-like field, $U_\mu$. The scalar
field, $\phi$ is governed by an additional action (cf. equation 25
of Bekenstein 2004). As a result, the scalar field $\phi$ tracks
the matter energy-momentum tensor distribution, satisfying an
equation,
 \beq\label{scalareq}
    \left[ {\tilde{\mu} \over 1-\tilde{\mu}}  \phi^{,\nu} \right]_{;\nu}
    = 4\pi G_K c^{-4} \left[g^{\mu\nu}+(1+{\rm e}^{-4\phi})U^\mu U^\nu \right]
    \tilde{T}_{\mu\nu},
 \eeq
where $\tilde{\mu}(\delta_\phi^2)$ is a
function of $\delta_\phi^2$ , which is defined by
 \beq
    \delta_\phi^2 \equiv {\phi^{,\nu} \phi_{,\nu} c^4\over a_0^2},
    \qquad \phi^{,\nu} \equiv (g^{\mu\nu} - U^\mu U^\nu) \phi_{,\mu}.
 \eeq
Here $\tilde{\mu}$ and $\delta_\phi^2$ are non-dynamical fields, i.e., they are
some fixed functions of the scalar field $\phi$ and the metric
$g_{\mu\nu}$.  As will become obvious $a_0$ can be identified with the acceleration
scale in MOND.

Near a galaxy the scalar field, $\phi$, is quasi-static, hence we
can neglect all time derivatives compared to spatial derivatives
so that
 \beq
    \delta_\phi^2 \approx {|\nablab \phi|^2 c^4\over a_0^2},
    \qquad \phi_{,\mu} \approx \phi^{,\mu} = (0, \nablab \phi).
 \eeq
Bekenstein argues that in the Einstein frame  
$g_{tt} \approx {\rm e}^{-2\phi} \tg_{tt} \approx - {\rm e}^{2 \Phi_N/c^2} c^2$, 
with $\Phi_N(\r) \equiv \Phi(\r)-\phi(\r) c^2$.
Substituting these, and a simplified energy momentum tensor
(equation \ref{enmomtensor}), into equation~(\ref{scalareq}) for
the scalar.  Drop the pressure term, 
time derivatives and higher order terms, which are all small
compared to the physical matter density term, $\tilde{\rho}$,
the equation now starts to resemble Poisson's equation,
 \beq\label{tevespoisson}
   2 \nablab \cdot
    \nablab  \Phi_N \approx 2 \nablab \cdot
    \left[ {\tilde{\mu} \over 1-\tilde{\mu}}   \nablab \phi c^2 \right] \approx 
   8\pi \widetilde{G} \tilde{\rho}, ~\widetilde{G} \equiv G_K {\rm e}^{-2\phi_c},
 \eeq
where we set $\phi \approx \phi_c$, which is the cosmological
average of $\phi$, and define $\widetilde{G}$
as the physical gravitational constant. Hence $\Phi_N(\r)$ is
basically the Newtonian gravitational potential, and is related to
the physical matter density by the Newtonian Poisson's equation.
The equation of motion in these potentials is
 \beq\label{teveseom}
    {d^2\r \over d\tilde{t}^2} =
    -\nablab\Phi(\r) = -\nablab \Phi_N(\r) - \nablab \phi(\r) c^2,
 \eeq
The above two equations link 
the gravitational potential $\Phi$, 
the scalar field $\phi$, the matter density $\tilde{\rho}$ and the motion $d\r/dt$
together.  Plus a certain choice of the free interpolation function,
{\it we fully specify the dynamics near a galaxy}.
As a first study we adopt a very simple choice of the free function with
\beq\label{mymu}\label{delta_mu}
\delta_\phi^2 = \cases {\tilde{\mu}^2(1-\tilde{\mu})^2, & \quad $0 \le \tilde{\mu} \le 1$, \cr
                0   & \quad otherwise .\cr}
\eeq
As will become obvious, this choice of $\delta_\phi^2$ is necessary to simplify the analytics of
lensing by an extended galaxy.  

To suit our studies, here we have chosen a set of notations
slightly different from Bekenstein.
We avoid using Bekenstein's $\mu$ completely, which does not have
the same role as Milgrom's $\mu$, which is a function of gravity
$g = |\nablab \Phi|$.  To distinguish from Milgrom's $\mu(g/a_0)$
as well, we use $\tilde{\mu}(\delta_\phi^2)$ to emphasise its
dependence on the scalar field strength, not the overall gravity.
\footnote{Our $G_K$ is related to Bekenstein's G and K by $G_K
\equiv G (1-K/2)^{-1}$, where $K$ is a small dimensionless
proportionality constant for the action of Bekenstein's vector
field.  Our $\tilde{\mu}$ is related to Bekenstein's scalar
$\sigma$ by ${\tilde{\mu} \over 1-\tilde{\mu}} \equiv  4\pi G_K
\sigma^2$. And $\delta_\phi^2 = \left[{4 \pi (1-K/2) \over
k}\right]^2 y/3$ where $y$ is Bekenstein's y-parameter. The $a_0$
is related to the Bekenstein parameters $k$ and $l$ by $a_0 =
{\sqrt{3k} \over 4 \pi \Xi l}$, and $\Xi=(1-K/2){\rm
e}^{-2\phi_c}$.}

\begin{figure}
 \psfig{figure=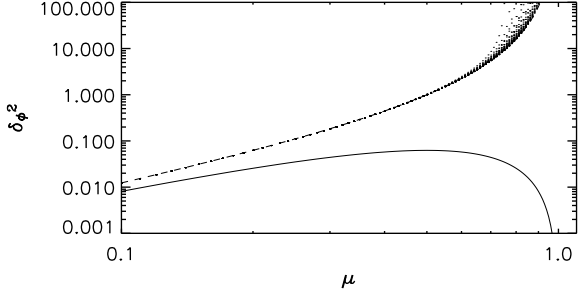,width=8cm,angle=0}
 \caption{shows the scalar field strength $\delta_\phi^2$ as function of $\tilde{\mu}$ in Bekenstein's toy models 
(region shaded by dots with $k$ and $K$ between $0$ and $10$).
Also shown is our principal model (solid line cf. eq.~\protect{\ref{mymu}}), and the smooth model 
(dashed line at the lower boundary Bekenstein models, cf. eq.~\protect{\ref{musmth}})
}\label{fig:mutilde}
\end{figure}

In our notations the toy model proposed by Bekenstein (his eq. 50) related 
$\delta_\phi^2$ and $\tilde{\mu}$ by
\beq\label{bekenmu}
\delta_\phi^2 = {\tilde{\mu}^2 \over (1- \tilde{\mu})^2}
\left[ {(1- {(1-K/2) k \over 8\pi} {\tilde{\mu} \over 1- \tilde{\mu}})^2 \over (1- {(1-K/2)k \over 4\pi}
{\tilde{\mu} \over 1- \tilde{\mu}}  ) } \right],
\eeq
where the parameters $k$ and $K$ are much less than unity.
Fig.~\ref{fig:mutilde} compares these Bekenstein's toy model with our toy model.
Our choice is different from Bekenstein's choice which allows
a $\delta_\phi^2$ with no upper limit and with a negative branch with subtle effects on cosmology.
Nevertheless, both choices preserve the asymptotic relation $\tilde{\mu} \rightarrow \delta_\phi$
in the weak gravity limit, which is essential for explaining galaxy rotation curves,
the main success of MOND.  We have also made lensing predictions for scalar field closely matching
Bekenstein's toy model, this is given in the Discussion section.

\section{Lensing and Potential in TeVeS}

\begin{figure}
\psfig{figure=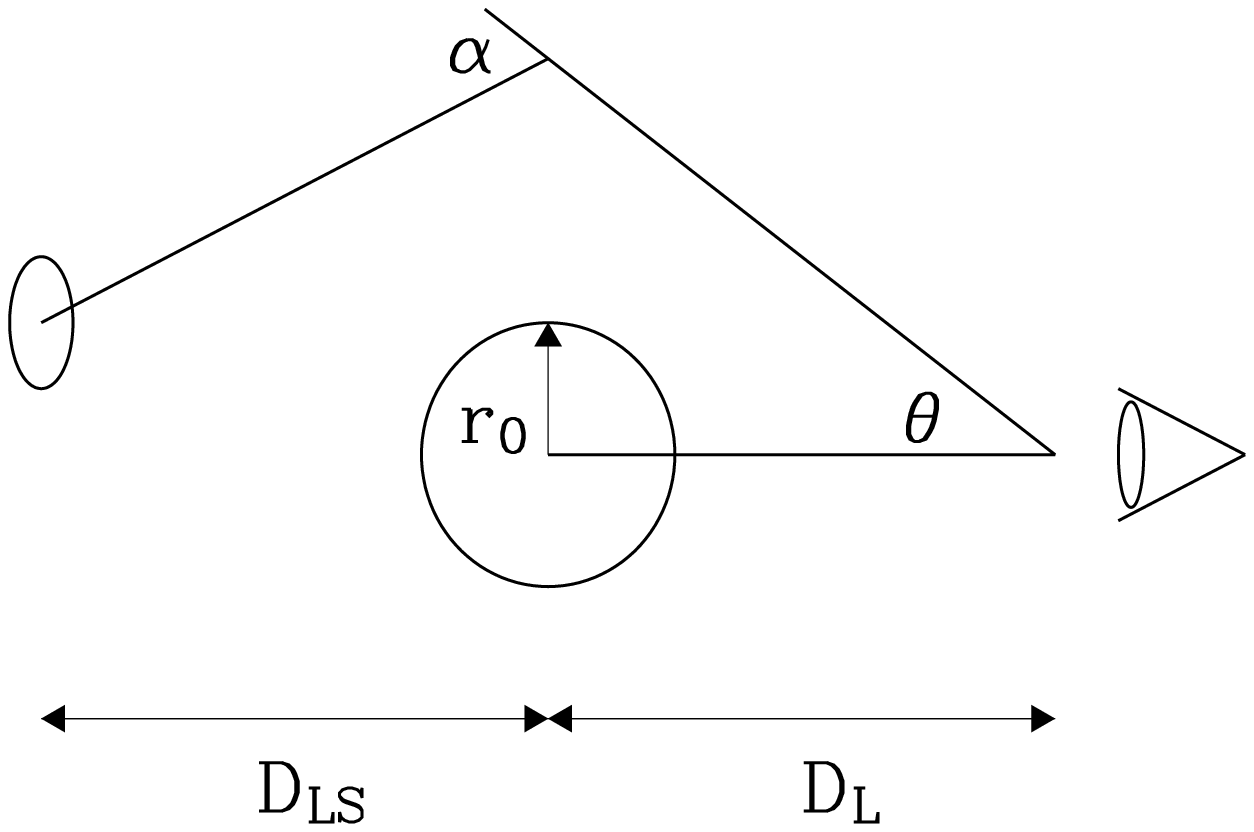,width=4cm}
\psfig{figure=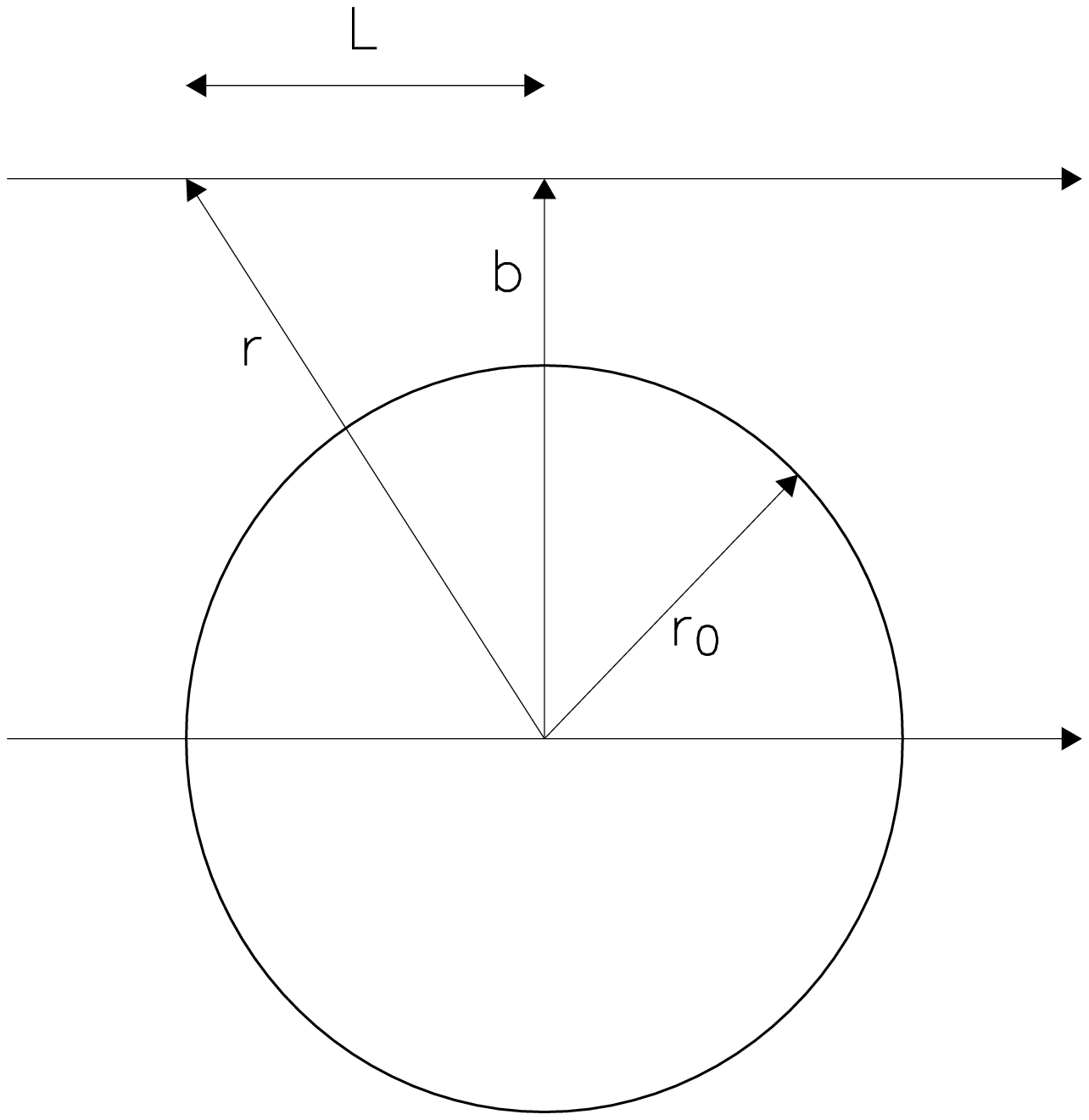,width=4cm,angle=0}
\caption{Schematics of the lens geometry considered in this
section. Top: Overall geometry. We indicate the angular diameter
distance from observer to lens, $D_l$ and lens to source,
$D_{ls}$, the observed image angle $\theta$, the bend angle
$\alpha$ and radius of Newtonian bubble $r_0$. Bottom: The
geometry of a light beam past a point lens, indicating the
position of a photon distance from the point-source, $r$, the
distance of closest approach, $b$ and the radius of the "Newtonian
bubble", $r_0=\sqrt{GM/a_0}$.}  \label{schem1}
\end{figure}

\subsection{Bending of Light Rays in Slightly Curved Space Time}

Light rays trace the null geodesics of the space time metric.
Lensing, or the trajectories of light rays in general, are
uniquely specified once the metric is given.  In this sense light
bending works exactly the same way in any relativistic theory as
in GR.

Near a quasi-static system like a galaxy, the physical space-time
is only slightly curved, and can be written in polar coordinates
as
 \ba
    -c^2 d\tau^2 & \approx &  {\rm e}^{2 \Phi/c^2} c^2 d\tilde{t}^2 + {\rm e}^{-2 \Phi/c^2} dl^2, \\
    dl^2&=&(dr^2+ r^2 d\theta^2+r^2\sin^2 \theta d\psi^2).
 \ea
where $|\Phi(\x)| \le c^2$ takes the meaning of a gravitational
potential in a rectangular coordinate $\x=(x_1,x_2,x_3)$ centred
on the galaxy, we note that a non-relativistic massive particle
moving in this metric follows the geodesic 
${d^2 \x \over d\tilde{t}^2} \approx - \nablab \Phi(\x)$,
which is the equation of motion in the non-relativistic limit.

Consider lensing by the potential $\Phi(\r)$ of a galaxy with a
geometry as in Fig.~\ref{schem1}.  An observed light ray travels a proper
distance $l_{os}=l_{ls}+l_{ol}$ from a source to the lens and then
to an observer.   
Hence it arrives after a time interval (seen in polar 
coordinates by an observer at rest with respect to the lens)
 \beq\label{geod}
       \tilde{t} 
\approx \int_{0}^{l_{os}} {\rm e}^{-2 \Phi({\bf r})/c^2} {dl \over c} 
\approx \int_{0}^{l_{os}} {dl \over c} - \int_{0}^{l_{os}}
       {2\Phi({\bf r}) \over c^2} {dl \over c},
 \eeq
where we used the fact that a light ray moving with a constant
speed $c$ inside follows the null geodesics $d\tau=0$.
As in an Einstein universe (c.f. Bartelmann \& Schneider 2001),
the arrival time contains a geometric term and a Shapiro time
delay term due to the $\Phi$ potential of a galaxy. Gravitational
time delay hence works as in GR, but with $\Phi$ instead of the
Newtonian gravitational potential, $\Phi_N(\r)$. Note that we
recover the GR-like factor of two in front of $\Phi$. Images will
be on extrema points of the light arrival surface, hence the
factor of two propagates to the deflection angle as well. Hence,
to a good approximation, lensing by galaxies in TeVeS behaves as
in GR apart from different interpretations of the gravitational
potential.

For example, assume a spherical lens with a potential $\Phi(r)$. A
light ray with an impact parameter $b$, 
we can reduce the eq. (109) of Bekenstein to
 \beq
    \alpha(b) \approx
    \int_{b}^{\infty}
    {d r \over c^2} {4 b \over \sqrt{r^2-b^2}}
    {d \Phi(r) \over d r },
\eeq
where we take the weak field and thin lens
approximation (i.e., we can drop higher order terms, assume the
lens is far from the observer, the source $l_{ls}\rightarrow
\infty$ and $l_{ol}\rightarrow \infty$, and approximating the closest approach 
$r_{\rm min}\approx b$).
Interestingly this is twice the bending angle predicted from
extrapolating non-relativistic dynamics (Qin et al. 1995).

In fact, gravitational lensing in TeVeS recovers many familiar
results of Einstein gravity even in non-spherical
geometries.  For example, an observer at redshift $z=0$ sees a
delay $\Delta t_{\rm obs}$
in the light arrival time due to a thin deflector at $z=z_l$
 \beq\label{tdr}
    {\Delta t_{\rm obs}({\bf R}) \over (1+z_l)}  \approx
    {D_{s} \over 2D_{l}D_{ls}}
    \left({\bf R}- {\bf R}_s \right)^2
    - \int_{-\infty}^{\infty} \!\!\! dl {2\Phi({\bf R},l) \over c^2},
 \eeq
as in GR for a weak-field thin lens, $\Phi/ c^2 \ll 1$. A light
ray penetrates the lens with a nearly straight line segment (within the thickness of the lens) 
with the 2-D coordinate, $\R=D_l \thetab$, perpendicular to the sky, where $D_l(z_l)=l_{ol}/(1+z_l)$
is the angular (diameter) distance of the lens at redshift $z_l$,
$D_s$ is the angular distances to the source, and $D_{ls}$ is the
angular distance from the lens to the source.
The usual lens equation can be obtained from the gradient of the
arrival time surface with respect to $\R$.  

Nevertheless, there are important differences between lensing in TeVeS and in GR.
These are mainly in the predicted metric for a given galaxy mass
distribution, and the predicted metric and distance-redshift
relation for the Hubble expansion, which we will come to.

There are, however, conceptual differences between TeVeS and MOND.
The modification of gravity in TeVeS is made through the factor 
$\tilde{\mu}(|\nablab \phi|^2c^4 a_0^{-2})$, hence depends on the 
scalar field gradient $|\nablab \phi|$, rather than $|\nablab \Phi|$
as in MOND.  Strictly speaking, $\nablab \phi$ and $\nablab
\Phi$ are generally not colinear except for special geometries,
hence the two descriptions are not generally identical.

In short Bekenstein's theory reduces to MOND in the
non-relativistic and spherical limit. Later in the paper, we will
work exclusively under the assumption that $\phi \sim \phi_c \sim
0$, or ${\rm e}^{\phi} \sim 1$. Hence we can drop the tilde sign
without confusion. We will consider spherical systems only, where
the magnitude of the gravitation field is given by equation
(\ref{teveseom}), and $a_0$ is given by equation (\ref{a0}).

\begin{figure}
\psfig{figure=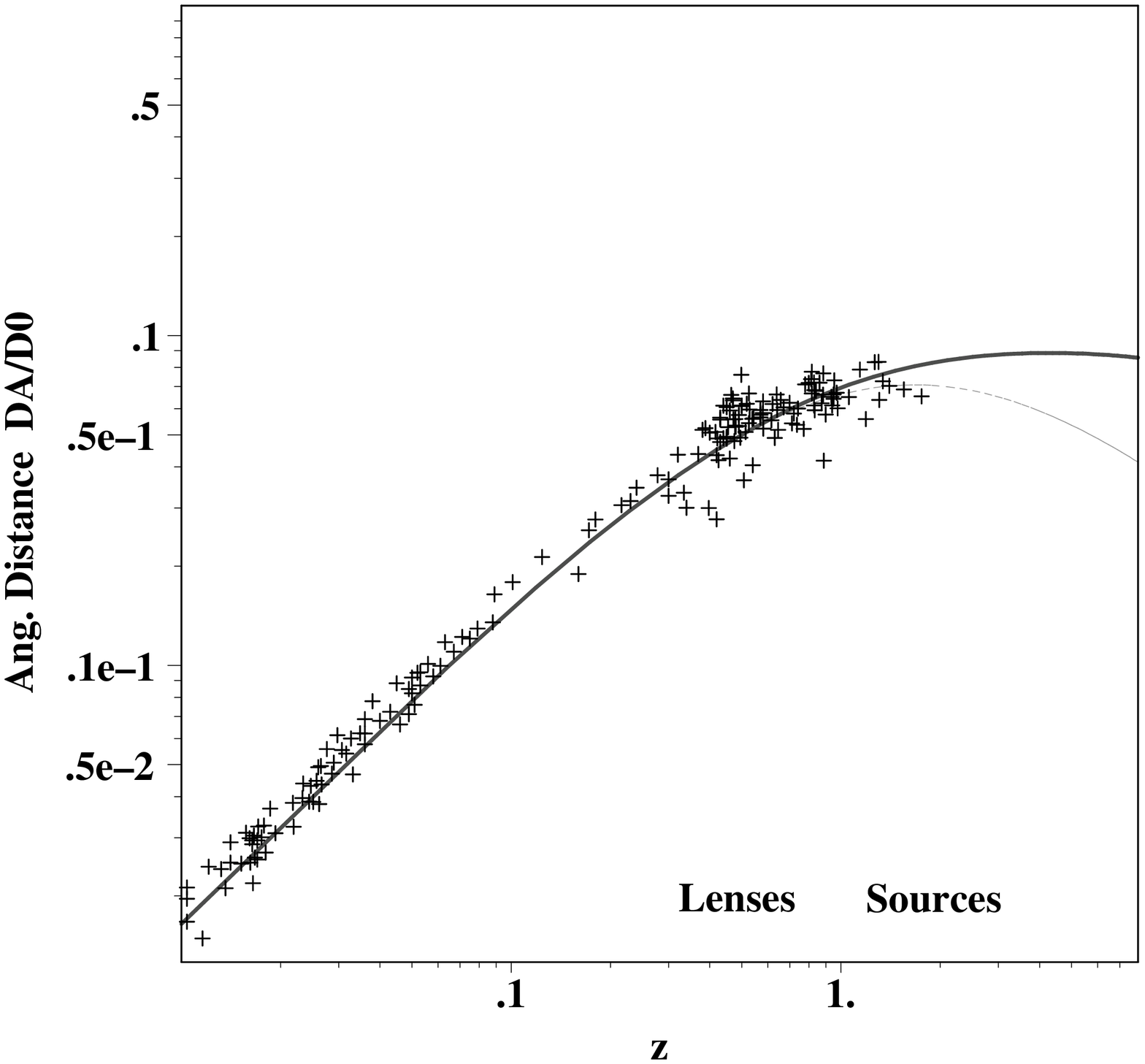,width=7cm,angle=0} 
\psfig{figure=sn2.ps,width=5cm,angle=-90} 
\psfig{figure=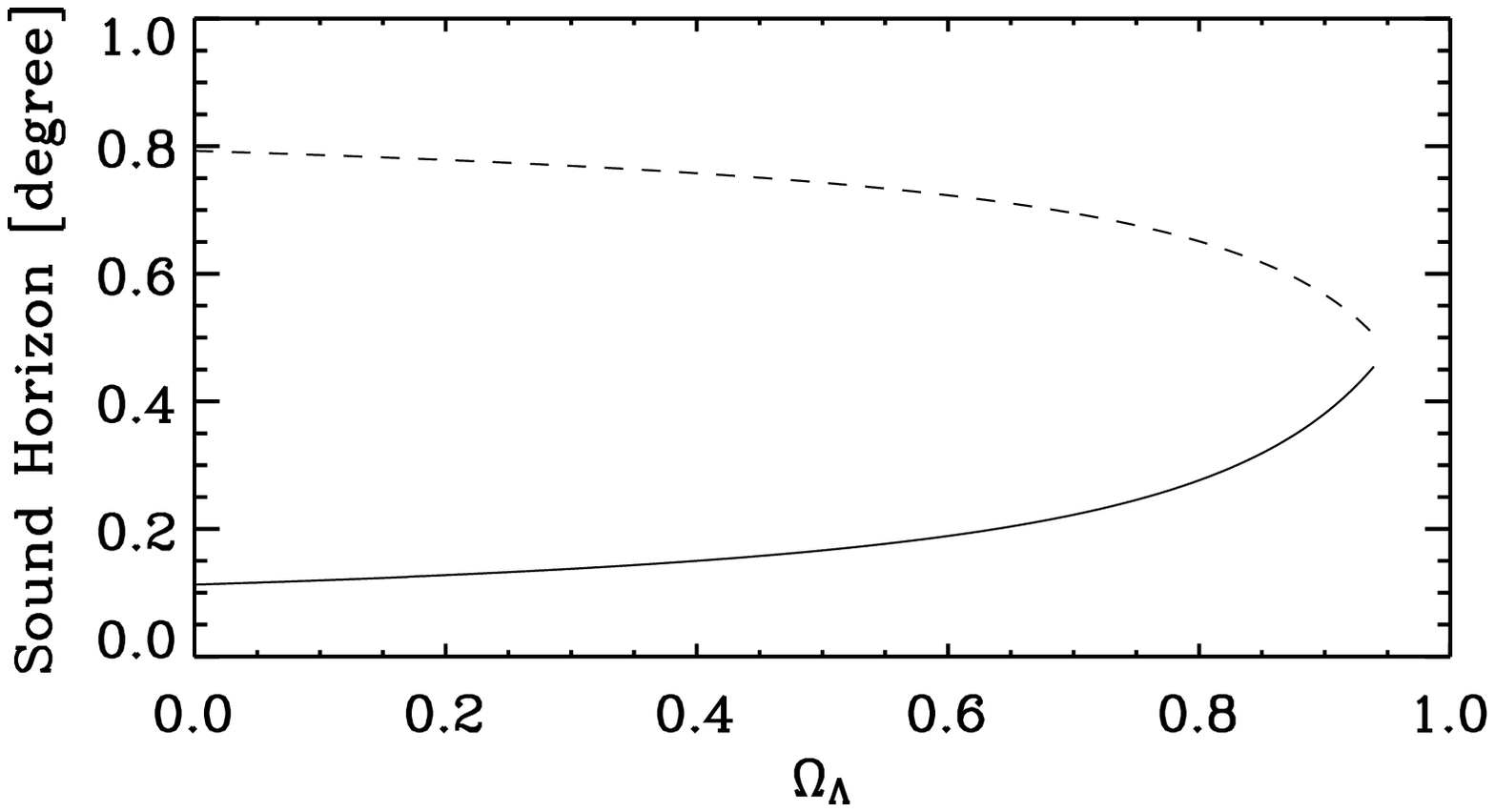,width=8cm,height=3.5cm,angle=0}
\caption{Upper panel: Angular diameter distance $D_A(z)$ (in units of the MOND critical distance 
$D_0 = c^2 a_0^{-1} \sim 6 cH_0^{-1}$) vs. redshift , $z$, in two cosmologies:
$(\Omega_m,\Omega_\Lambda)=(0.04,0.46)$ ("minimal-matter" cosmology, thick solid) and
$(0.25,0.75)$ (LCDM, thin dashed).  The data are converted to
angular diameter distances by multiplying by a factor $(1+z)^{-2}$
and assuming $H_0=70\,{\rm km}s^{-1}{\rm Mpc}^{-1}$.  We also
label typical lenses at redshifts $z=0.1$ -- $1$, and typical
quasar sources at redshifts $z=1$ -- $3$ (as labelled).
Note ${D_A(z) \over D_0} \ll 1$ {\it at all redshifts} due to the low value 
of $a_0$, which means that cosmological objects have too small angular diameter
distances for any prominent deep-MOND effect in lensing, i.e., most of 
the bending is done in the Newtonian bubble of Fig.~\protect{\ref{schem1}}).  
Middle panel: The minimal-matter cosmology (TeVeS) fit to 
high-z SN Ia luminosity distance
modulus is slightly poorer than the $\Lambda$CDM flat cosmology fit, although only by
$\Delta \chi^2=3.5$ or $1.9\sigma$; error bars indicate the $\pm 1\sigma$ limit
on the estimated $\Omega_\Lambda$.  
Bottom panel: shows for models of different $\Omega_\Lambda$ 
the sound horizon at $z=1000$ (in degrees) 
in the above "minimal-matter" open cosmology and CDM-based closed cosmology.
}\label{fig:SN}\label{fig:DAz}
\end{figure}

\section{Angular Diameter Distances for Lensing in a simplistic minimal-matter open cosmology}

To describe gravitational lensing, it is essential to know how the
background cosmology behaves in Bekenstein's TeVeS.  
Lensing requires knowing the luminosity distance $D_L(z)$ 
and the angular diameter distance $D_A$ in a universe as functions of redshift.
These can be predicted generally by
 \be
        {D_L (z)\over (1+z)^2}= D_A(z) = {c \over (1+z)}
        { \sinh\left[ \Omega_K^{1/2}
        \int_0^{z} {dz \over H(z)}\right] \over \Omega_K^{1/2}}.
 \ee
where $\Omega_K$ is the usual curvature density parameter
and $H(z)$ is the Hubble expansion rate.  
To develop the cosmological model,
we assume an isotropic and homogeneous physical metric in TeVeS
(cf. equations 117 and 118 of Bekenstein);
 \beq
    -c^2 d\tau^2  =  -c^2 d\tilde{t}^2 +
    \tilde{a}(\tilde{t})^2 \left[ d\chi^2 +
    \sinh\chi^2 \left( d\theta^2+ \sin\theta^2 d\psi^2 \right)\right]
 \eeq
for an open universe with the expansion factor $\tilde{a}$ and the cosmological time $\tilde{t}$.
   
We are interested in lensing in a reasonable cosmological model in TeVeS context,
which should be in accord with Supernovae (SNIa) data.  
There are many degrees of freedom in TeVeS, which is beyond the exploration of this paper.
One way is to consider a "mininal cosmology" in TeVeS context
with just a low density baryonic universe plus some amount of cosmological constant model
such that to fit SNe data.  
As argued by Bekenstein, the scalar field contributes negligibly
to the Hubble expansion, with a ratio $O(k)$ compared to the matter
contribution, and $k$ is small. Hence the minimal-matter cosmological model 
is largely the same as in the case of GR, and for latter
discussion we set the energy density of the scalar field
$\rho_\phi=0$ (c.f. Section VII, Bekenstein 2004).
Under this {\it crude approximation}, we find that the Hubble parameter $H(z)$ is given by
 \be
        {\dot{\tilde{a}}^2 \over \tilde{a}^2} = H^2(z) \approx H_0^2
        \left(\Omega_m (1+z)^3 + \Omega_\Lambda + \Omega_K (1+z)^2 \right),
 \ee
where $\Omega_K \approx 1-\Omega_\Lambda - \Omega_m$.
At low redshift the vacuum energy is
important and dominating.  Choosing an appropriate vacuum energy, 
it is conceivable that one can make a
TeVeS baryonic universe which expands at virtually the same rate
as the CDM model. For example, 
we find a reasonable fit is an open cosmology with 
$\Omega_\Lambda \sim 0.46$, $\Omega_m \sim 0.03$, and $H_0 \sim 70$km/s/Mpc.
It is possible to
fit the luminosity distances of high-z SNe about as well as CDM up
to a redshift of 1 -- 2.  Fig.~\ref{fig:DAz} show the angular diameter 
distance as a function of redshift, for a
best-fit minimal-matter cosmology and a standard $\Lambda$CDM model.  Clearly
both models are consistent with the high-z SNe distance moduli data
set; the minimal-matter cosmology fit is slightly poorer than the $\Lambda$CDM fit, but
only by 1.9$\sigma$ and is therefore admissible.  This open cosmology has 
problems at very high redshift, e.g., the last scattering sound horizon will be too small.
This undesirable feature might be {\it an artifact of the our simple cosmology, rather than
intrinsic to TeVeS}; e.g., 
allowing for $2$eV neutrino Dark Matter to co-exist with baryonic matter in MOND
would soften the problems (Sanders 2003).  More detailed analysis is clearly needed.
Despite its limitations 
our simplistic minimal-matter cosmology is sufficient for assigning lensing and source 
distances in redshift $z=1-3$.

\begin{table}
\caption[ ]{Lookup-table for our notations for lensing in TeVeS}
\begin{tabular}{l|l}
Parameters & Meaning \\
\hline
$a_0$ &  Threshold for weak acceleration regime\\
$g$   & Gravity strength $|\grad \Phi|$ in TeVeS \\
$g_N$ & Newtonian gravity strength $|\grad \Phi_N|$ \\
$\mu(g)$ or $\tilde{\mu}(\delta_\phi^2)$ & Interpolation function\\
$\grad \phi c^2 =a_0\sqrt{|\delta_\phi|}$ & Scalar field strength in quasi-static galaxies \\
$D_l$ &  Angular diameter distance to lens \\
$D_s$ & Angular diameter distance to source \\
$D_{ls}$ & Angular distance from lens to source \\
$D_0=c^2/a_0$ & Distance scale of weak acceleration \\
$\xi={\eta^2 \over 4}={D_lD_{ls} \over D_s D_0}$ & Rescaled
lens/source effective distance \\
$r_0=\theta_0 D_l=\sqrt{GM \over a_0}$  & Newtonian bubble radius and angular radius \\
$r_h=\theta_h D_l$   & Hernquist lens scale length and angular scale\\
$\theta=b/D_l$ & Angular size of closest approach to lens \\
$\alpha$ & Deflection angle \\
$\baralpha=\frac{D_{ls}} {D_s} \alpha $ & Reduced deflection angle \\
$\theta_c$ & Critical (Einstein ring) angular radius \\
$y={\theta \over \theta_0}={b \over r_0}$ & Rescaled angular radius
of the lens \\
$y_c=\theta_c/\theta_0$ & Rescaled angular critical radius \\
\end{tabular}
\end{table}

\section{Modelling a spherical lens and a point lens}

Having set out the basics of gravitational lensing and the
cosmological model in the relativistic
TeVeS theory, we now consider analytic solutions to simple
spherical lensing models.  
Equation (\ref{tevespoisson}) reduces to the MONDian equations for
$\Phi(\r)$,
 \beq\label{sphericalteves}
    \nablab \cdot \left[ \tilde{\mu} \nablab \Phi \right] =
    4\pi \tilde{G} \tilde{\rho}, \qquad \tilde{\mu} \nablab \Phi =
    \nablab \Phi_N,
 \eeq
if we assume the two gradients $\nablab \Phi$ and $\nablab \phi$
are parallel (as in a sphere or far from a non-spherical body) or
if the gradient $\nablab \phi$ is negligible (as in strong
gravity).

The simplest case of lensing is by a point-like mass distribution.
Chiu et al. (2005) have worked out the deflection and time delay in
the point lens case rigorously for TeVeS.  We follow the more GR-like
formulation of Bekenstein (2004), neglecting high order terms.  The
results of the two approaches are essentially the same when dealing with
a galaxy potential.
Qin et al. (1995) and Mortlock \& Turner (2001) have
previously worked out deflections and amplification by a point lens.
Here we expand these earlier works by predicting divergence, shear and
critical line for a general spherical lens.
We will also fit the observed lenses later in the paper.
A list of notations are given in Table I.

Consider a geometry as illustrated by Figure~\ref{schem1},  
where a spherical lens at redshift $z_l$ bends the ray from 
a source at redshift $z_s$.  
Following the convention in gravitational lensing to work with
angles projected on the sky, we let the source be offset from the lens by an angle
$\theta_s$ and form an image at angle $\theta$, which is related to the physical
length $b$ (for the closest approach) by
 \beq
    \theta = {b \over D_l}.
 \eeq
The spherical symmetry of the lens means that 
the line of sight to the lens, source and images lie in one plane.  
Many of the familiar results from gravitational lensing in Einstein gravity transfer
directly over to TeVeS. 
Taking the derivative of eq.~(\ref{tdr}) with respect to $R$, and requiring images 
to form at extreme points, we find the lens equation 
 \beq\label{lenseq}
    \theta - \theta_s = \baralpha \equiv {D_{ls} \over D_s} \alpha, ~
    \alpha(b) = \int_{b}^{\infty} {d r \over c^2} {4 b \over \sqrt{r^2-b^2}} {d \Phi(r) \over d r}.
 \eeq
Taking one more derivative to $\theta_s$, we find that the amplification $A$ is given by
\beq
A^{-1} = {\theta_s d\theta_s \over \theta d\theta} = 
(1-\kappa-\gamma) (1-\kappa+\gamma).
\eeq
Here $\kappa$ and $\gamma$ are the convergence field and the shear field, still given by
 \be
    \kappa = \frac{1}{2\theta} \frac{\de}{\de \theta} \theta^2 \barkappa, 
~ \gamma = | \barkappa - \kappa|, 
~\barkappa = {\baralpha \over \theta \barkappa} = \frac{2}{\theta^2} \int_0^{\theta}  d\theta
    (\theta \, \kappa). 
 \ee
where $\barkappa$ is the mean convergence within a circular radius.

To offer more insight, we can re-write the deflection angle (cf. eq.~\ref{lenseq}) in terms of
the photon position along an unperturbed path (the Born
approximation) as
 \beq
    \alpha \approx
    \int_{-\infty}^{\infty} {d l \over c^2} 2 g_{\perp}(r),
 \qquad r=\sqrt{l^2+b^2},
 \label{alpdef}
 \eeq
where $l$ is the distance long the light path and
 \beq
    g_{\perp}(r) = g(r) {b \over r},\hspace{0.5cm}g(r)={d \Phi(r) \over d r },
    \label{gperpdef}
 \ee
is the gravitational force acting transversely to the direction of
the photon motion at a distance $r$ from the source. 

To simplify notations later on, 
it is helpful to scale the distances to a TeVeS distance scale $D_0$ defined by
 \beq
    D_0 = {c^2 \over a_0} \approx {6 c \over H_0} \sim 25 {\rm Gpc}.
 \eeq
Heuristically, this gives a characteristic distance in TeVeS at
which a particle accelerated with $a=a_0$ would reach the speed of
light (ignoring relativistic effects). 
Define $\eta$ as a dimensionless number with 
\beq\label{eta}
\eta \equiv \sqrt{4\xi}, ~  \xi \equiv {D \over D_0}, ~D={D_l D_{ls} \over D_s}.
\eeq
The quantity $\eta$ or $\xi$ characterises the geometry of the lens system, independently of the lens mass.
For a point lens of mass $M$
the meaning of $\eta$ can also be recognised from the fact that
$\eta \theta_0=\sqrt{ 4 G M D_{ls} \over c^2 D_s D_l}
= 0.8\arcsec\sqrt{{M \over 10^{11}\msun}{1 {\rm Gpc} \over D_l} {D_{ls} \over D_s}}$
is the conventional Einstein radius in the GR limit; this does not hold in TeVeS.
Generally $\eta$ or $\xi$ represent the lens geometry and are
independent of the lens mass $M$.

With our choice of $\delta_\phi$ as a function of $\tilde{\mu}$
relation (equation \ref{delta_mu}) we can further simplify the
gravity inside a spherical system.  We find
 \beq
 \label{tevesstep}
        g(r)=\nabla \Phi = \cases{\sqrt{a_0 g_N } &
        $ g_N(r) \le a_0$,\cr
               g_N & \quad  otherwise.\cr}
\eeq
which is a special case of eq.~(\ref{mondstep}) of Milgrom's theory.
Here the Newtonian gravity $g_N(r)$ and potential $\Phi_N(r)$ are given by
 \beq\label{gnewton}
        g_N(r) = {G M(r) \over r^2} = {d \Phi_N(r) \over dr},
 \eeq
where $M(r)$ is the mass enclosed inside a spherical physical (proper) radius $r$.
In TeVeS, the gravitational field $g$ for a spherically-symmetric
source is made up of two contributions, $\grad \phi$ from the
scalar field and $g_N$ from the Newtonian gravity. For our choice
of the free function we find that gravity in TeVeS is related to
Newtonian gravity by
 \be\label{gscalar}
    \grad \phi c^2 = g- g_N = \cases {  0,  \quad &  $g_N > a_0$,\cr
             \sqrt{g_N a_0}-g_N \quad & $g_N < a_0$.\cr}
 \ee

\subsection{Lensing by a Point mass}

Around a point-mass the gravity (cf. eq.~\ref{tevesstep})
 \be
    g= \cases {  a_0 (r_0/r)^2,  \quad &  $r < r_0$,\cr
                 a_0 (r_0/r) \quad & $r> r_0$,\cr}
    \label{pointg}
 \ee
where we have defined a transition radius $r_0$ by $g(r=r_0)=a_0$,
so that inside a Newtonian bubble, $r<r_0$, around the point-lens we
have strong Newtonian gravity where $g\propto 1/r^2$, and outside
we have weak TeVeS gravity where $g\propto 1/r$. 
At $r=r_0$ we have
 \bey
    r_0 &=& \sqrt{G M \over a_0} = {G M \over v_0^2} =10\kpc \sqrt{M \over 10^{11}\msun} ,\\
     v_0 &=& ( G M a_0 )^{1/4} = 200\kms \left({M \over 10^{11}\msun}\right)^{1 \over 4} ,
 \eey
is the circular velocity in the weak gravity regime outside the
Newtonian bubble.

Substituting the TeVeS gravity $g(r)$ (eq.~\ref{pointg}) into eq.~(\ref{alpdef}), we find the deflection angle
 \ba
    \alpha &=& {4 G M \over c^2 b} \sqrt{1-\left({b \over r_0}\right)^2} +
    {4 v_0^2 \over c^2} \sin^{-1}\!{b \over r_0}, \qquad b<r_0,\\
           &=& {2 \pi v_0^2 \over c^2},  \qquad \hspace{4.0cm} b \ge
           r_0.
           \label{mondalp}
 \ea
This result is straightforward to understand: for a light path
with a large enough impact parameter $b$, we have $r > r_0$
everywhere along the line of sight so that the TeVeS gravity looks
like that of an isothermal halo of circular velocity $v_0$, and
hence we recover the result of a constant deflection as in GR for
isothermal halos. In the other limit where $b \ll r_0$, the line
of sight will go through the strong gravity, Newtonian regime and
the deflection approaches that of a point mass in GR,
 \be
    \alpha_{\rm GR} ={4 v_0^2 b_0 \over c^2 b}= {4 G M \over c^2 b}.
 \ee
 This term also appears as the leading term for small distances
 from the source in equation (\ref{mondalp}). The extra terms in
 equation (\ref{mondalp}) are due to modifications as the light beam
 passes through the weaker, MOND-gravity regime, when $r>r_0$.

\begin{figure}
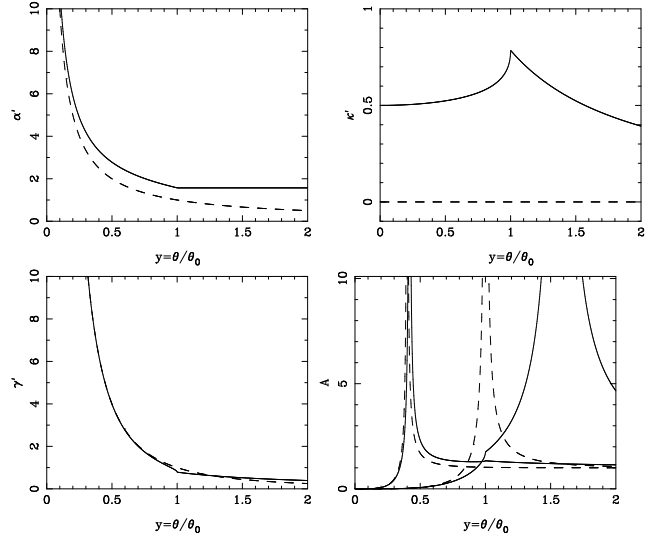

 \psfig{figure=alp.ps,width=4cm,angle=0}
 \psfig{figure=kap.ps,width=4cm,angle=0} 
 \psfig{figure=gam.ps,width=4cm,angle=0} 
 \psfig{figure=mag.ps,width=4cm,angle=0}
\caption{Shows dimensionless functions of the rescaled impact parameter 
$y=\theta/\theta_0$ in TeVeS (solid lines) and in GR (dashed lines): 
(a): deflection angle $\baralpha/(\theta_0 \eta^2)$; beyond the
 Newtonian bubble of angular radius $\theta_0$, the deflection is a constant. 
(b): convergence $\kappa/\eta^2$; note the cusp at the
 transition between weak and strong MOND regimes; also $\kappa=0$ in GR for a point lens.
(c): shear   $\gamma/\eta^2$; the signal increases significantly
 within the Newtonian bubble.  Panel (d) we shows two magnification profiles for a 
 point-mass source in TeVeS (solid) and in GR (dashed), for values of the dimensionless 
 MOND-lens geometric parameter, $\eta=2\sqrt{\xi}=0.4$ and $=1$. } \label{mag}
 \label{alpkapgam}
\end{figure}

It is convenient to work with dimensionless quantities to find
universal relations within TeVeS. 
Define $\theta_0$ as the angular size of the Newtonian bubble
with physical radius $r_0$.  
 \be
    \theta_0 ={r_0 \over D_l} = \sqrt{\frac{GM}{a_0}} \frac{1}{D_l}
    = 2 \arcsec \left({M \over 10^{11}M_\odot} \right)^{1 \over 2}
    \left( \frac{D_l(z)}{1 {\rm Gpc}}\right)^{-1}.
 \ee
We can then express the image angle $\theta$ 
in terms of the dimensionless angle $y$ where
 \be
    y \equiv \frac{\theta}{\theta_0} = \frac{b}{r_0},
 \ee
and find the deflection angle satisfies
 \be\label{alpeta}
    {\baralpha \over \theta_0 \eta^2} = \cases{ \sqrt{y^{-2}-1} + \sin^{-1}\!y, & $y<1$,\cr
             \frac{\pi}{2},  & $y \ge 1$,\cr}
 \ee
where $\eta$ is given in eq.~\ref{eta}.
We can also transform the convergence and shear
into dimensionless quantities.
 \be\label{kapeta}
    \frac{\kappa}{\eta^2} = 
\cases{ \frac{1}{2} \left( \delta_D(\y)
    +\frac{\sin^{-1}y}{y}\right), & $y<1$,\cr
                              \frac{\pi}{4 y}, & $y \ge 1$, \cr}
 \ee
 \be\label{gameta}
    \frac{\gamma}{\eta^2} =\cases{ \left| \frac{\sqrt{1-y^2}}{y^2} +
    \frac{\sin^{-1}y}{2y}\right|, & $y<1$,\cr
             \frac{\pi}{4 y}, & $y \ge 1$.\cr}
 \ee
The amplification $A=|(1-\kappa-\gamma)(1-\kappa+\gamma)|^{-1}$.
These dimensionless results are plotted in 
Figures \ref{alpkapgam}.
We see that both the deflection angle
and the shear decrease from the central point, just as for the point-like lens
in Einstein gravity. But beyond the MOND-angle, $\theta=\theta_0$,
the deflection angle is a constant, and the shear falls slowly as $y^{-1}$, 
just as it is for an isothermal sphere in Einstein gravity. This is perhaps not too
surprising, since the aim of MOND was to mimic the rotation curve
of a dark matter-dominated isothermal sphere with a point-like
MOND source. As we have modified gravity in the same way for
gravitational lensing, we find a similar result.

The shape of the convergence field, $\kappa/\eta^2$, in Figure \ref{alpkapgam}
can also be simply understood. For angles less than the
MOND-angle, $\theta_0$, a light ray will pass for most of its path
through the Universe in the weak MOND regime. We assume there is
some large-scale cut-off to this and we return to the standard
cosmological model on very large scales. In this weak MOND-regime
the light bundle will experience a convergence. Then, when the
light ray passes through the Newtonian bubble, it experiences no
further convergence, except for those rays that pass through the
central point-mass. At the MOND-radius, $\theta=\theta_0$, there
is a cusp as light rays intersect the edge of the Newtonian
bubble. This cusp is a result of the discontinuity in $g$, given
by equation (\ref{pointg}), and other versions of TeVeS will give
smoother transitions. For light rays that pass further away from
the point-lens and never intersect the Newtonian bubble, the
convergence falls off as $y^{-2}$, as for an isothermal sphere in
Einstein gravity.  The magnification pattern depends on the lens 
distance parameter $\eta$ or $\xi$, and is substantially different from 
GR prediction if $\eta$ is not very small.


Further insight into the behaviour of these results can be found
by taking the limit $y \ll 1$. In this limit we find
 \ba\label{alphapt}
 {\baralpha \over \theta_0 \eta^2} &\approx & \frac{1}{y}\left(1+\frac{1}{2}y^2 \right),\\
 \frac{\kappa}{\eta^2}  &\approx & \frac{1}{2} \left( \delta_D(\y) +1 +
                \frac{1}{6}y^2\right),\\
 \frac{\kappa}{\eta^2}  &\approx & \frac{1}{y^2} \left(
            1-\frac{1}{24}y^4\right),\\
 A &\approx & \left| 1- \eta^2 + {\eta^4 \over 3} - {\eta^4 \over y^4} -
 \left({\eta^2 \over 6}-{2 \eta^4 \over 15}\right)y^2 \right|^{-1},
 \ea
which show the first-order corrections to lensing by a point-lens
in Einstein gravity. In physical coordinates the convergence field is
  \be
    \kappa \approx {2 G M D_{ls}\over D_s D_l c^2} \delta_D(\thetab) + 2\xi, \qquad \xi={D_{ls} D_l \over D_s D_0}
  \ee
which reduces to the usual result for a point lens in Einstein
gravity when $D_0 \rightarrow \infty$. Hence the first-order
correction to the usual Einstein gravity results for gravitational
lensing for TeVeS is an added constant of $2\xi={\eta^2 \over 2}$.

\subsection{Critical line around a point lens}

In general, the critical radius is where the magnification diverges.
For a spherical lens, $A \rightarrow \infty$ when $\theta_s=0$.
The critical radius, $\theta=\theta_c$, can be solved by
setting $\theta_s=\theta-\baralpha=0$ in the lens equation (cf. eq.~\ref{alpeta}).
We find that the rescaled critical radius
$y_c ={\theta_c \over \theta_0}$ is a function $\xi$, 
satisfying
 \ba\label{xipt}
    \xi={\eta^2 \over 4} &=& {y_c \over 4 \sqrt{y_c^{-2}-1} + 4 \arcsin y_c } \qquad y_c<1,\\
        &=& {y_c \over 2\pi} \qquad {\rm otherwise}
 \ea
which could be inverted approximately by interpolating the
asymptotic relations,
 \beq\label{eq:MLpt}
        y_c(\xi) = \left({G M \over a_0}\right)^{-1/2} D_l \theta_c  \sim \sqrt{4\xi+ 4\pi^2 \xi^2}.
 \eeq
Note that we avoid the term "Einstein radius", which normally means an angle $\sqrt{4 GM
D_{ls}/D_s D_l c^2}$ in GR.  The critical radius
$\theta_c =\theta_0 y_c(\xi)$ in TeVeS is also proportional to
$\sqrt{M}$, but depends on the distances through $\xi$ in a more sophisticated way.

In the MOND limit for $y_c \ll 1$ and $\xi \ll 1$ we find the
critical lines lie at
 \be
       y_c = {\theta_c \over \theta_0} = 2 \sqrt{\xi} (1+\xi).
 \ee

In the limit that $y_c \gg 1$, the TeVeS point-mass looks like an
isothermal sphere in Einstein gravity, with a constant deflection
angle.  Here the magnification, convergence and shear fields are given by
 \be
    A = |1-2 \kappa|^{-1}, \qquad \kappa=\gamma = \frac{\pi \xi \theta_0}{\theta}.
 \ee
Any critical lines occur when $\kappa=1/2$ at a radius of
 \be
    y_c = 2\pi \xi,     \,\,\,\,\,\,\,{\rm or}
    \,\,\,\,\,\,\,      \theta_c = 2 \pi \xi \theta_0=0.56\arcsec
\sqrt{M \over 10^{11}\msun}{D_{ls} \over D_s}.
 \label{mondcrit}
 \ee

\begin{figure}
 \psfig{figure=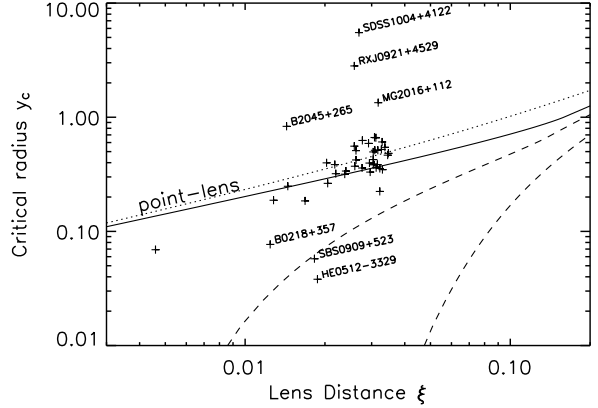,width=8.5cm,angle=0}
 \caption{Shows the critical radius $y_c=D_l \theta_c/r_0$ (rescaled by Newtonian bubble $r_0=\sqrt{GM/a_0}$),
 as a function of the lens effective distance parameter,
 $\xi=\eta^2/4=D_lD_{ls}/D_s/D_0$ (rescaled by the MOND distance $D_0=c^2/a_0$), 
 where $a_0=10^{-8}{\rm cm}{\rm s}^{-2}$.  
 Crosses indicate 44 lens systems 
 with data from the CASTLES lensing survey, including those in Table 2 and Table 3.
 The curves are for a point lens (solid), and a Hernquist profile lens with a scale length
 $r_h =r_0/3$ (dashed), and $r_0$ (lower dashed).  
}
 \label{ycdata}
\end{figure}

In Figure \ref{ycdata} (solid line for point mass lens) we have
plotted the dimensionless position of the critical lines,
$y_c=\theta_c/\theta_0$, as a function of the lens geometric
factor, $\xi = {D_l D_{ls} \over D_s D_0}$. At
small values of the critical line, and small values of $\xi$, we are
in the strong-gravity regime and so we expect to recover the results
of Einstein gravity, where $y_c
\propto \sqrt{\xi}$. In the weak-gravity MOND regime we expect a
transition to equation (\ref{mondcrit}), where $y_c \propto \xi$.
Hence, we can use data from the observed measurement of
critical lines around galaxies with known baryon content to test this
prediction of relativistic-MOND theory: basically we compare 
the expected baryonic mass of a lens of observed luminosity $L$
with the gravitational mass $M$ inferred from the observed Einstein ring, 
which satisfies eq.~(\ref{eq:MLpt}).
We carry out this procedure in \S 7.1.

\section{Lens model for baryonic Hernquist profile}

Early type galaxies are fairly well described by
a nearly round distribution of light with a de Vaucouleur radial profile in projection. 
In this section we shall extend our results from a point-like lens
to an extended, Hernquist-profile lens model.  The latter has a density
profile $\rho(r)$ and enclosed mass profile $M(r)$ given by
 \beq
    \rho(r)={M r_h \over 2\pi r (r+r_h)^3}, \qquad M(r)={r \over r+r_h}M,
 \eeq
where $M$ is the total mass and $r_h$ is the core scale length. This Hernquist
model is a good fit to the de Vaucouleur profile of elliptical galaxies, 
and is frequently used for quantifying these profiles 
(e.g. Kochanek et al 2000, Kochanek 2003). Since in a TeVeS
context the mass must follow light, a Hernquist profile should
provide an accurate lens model.

\subsection{Gravity and scalar field}

The Hernquist model has a Newtonian gravity 
\beq
g_N(r) = {G M(r) \over r^2} = {G M \over (r+r_h)^2}.
\eeq
Hence in TeVeS the gravity is given by
\bey
g(r) &=& {G M \over (r+r_h)^2}, \qquad r < r_0-r_h \nn &=& {v_0^2
\over r+r_h}, \qquad r> r_0-r_h
    \label{herngrav}
\eey
where $r=r_0-r_h$ is now the radius of the Newtonian bubble between
strong and weak gravity.   Note that the gravity reaches a finite maximum
at the centre of a Hernquist profile.  We define
 \be
        \theta_h= \frac{r_h}{D_L}
 \ee
as the angular size of the core length scale, $r_h$, projected on the
sky.  A point mass model is obtained in case that $\theta_h \rightarrow 0$.

The scalar field gradient around a Hernquist profile galaxy 
is given by (cf. eqs.~\ref{delta_mu},~\ref{gscalar}),
 \be\label{hernscalar}
 {\grad\phi c^2 \over a_0 }= \tilde{\mu}(1-\tilde{\mu}) = 
 \cases{ 0, & $r < r_0-r_h$,\cr
            {r_0  \over r+r_h} \left[1- {r_0 \over r+r_h}\right], & otherwise. \cr}
 \ee
Here the parameter $\tilde{\mu}=\min(1,{r_0 \over r+r_h})$, hence satisfying 
\beq\label{mujump}
\tilde{\mu} = \min(1,{g \over a_0})= {1 \over 2} \pm \sqrt{{1 \over 4} -{\grad\phi c^2 \over a_0}}.
\eeq
Integrating the above from $r$ to the infinity,  we find the scalar field 
 \be
 \phi(r) = \cases{ 0, & $r < r_0-r_h$,\cr
           {v_0^2 \over c^2} \ln{r+r_h \over r_0} + {G M \over c^2 (r+r_h)}-{G M \over c^2 r_0} , &
           otherwise.\cr}
 \ee
Hence $\phi$ contributes only beyond the Newtonian bubble. 
\begin{figure}
 \psfig{figure=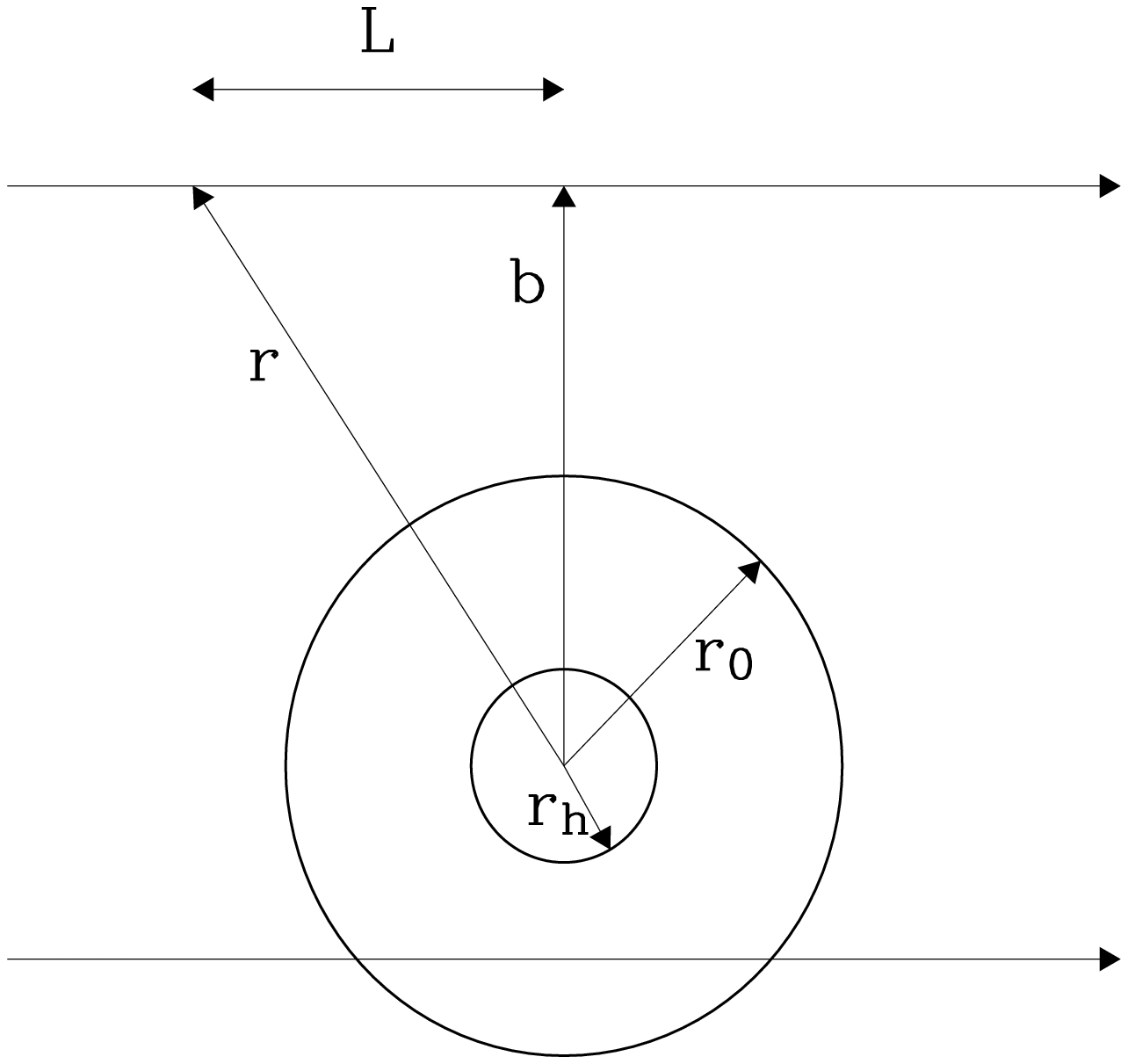,width=4cm,angle=0}
 \psfig{figure=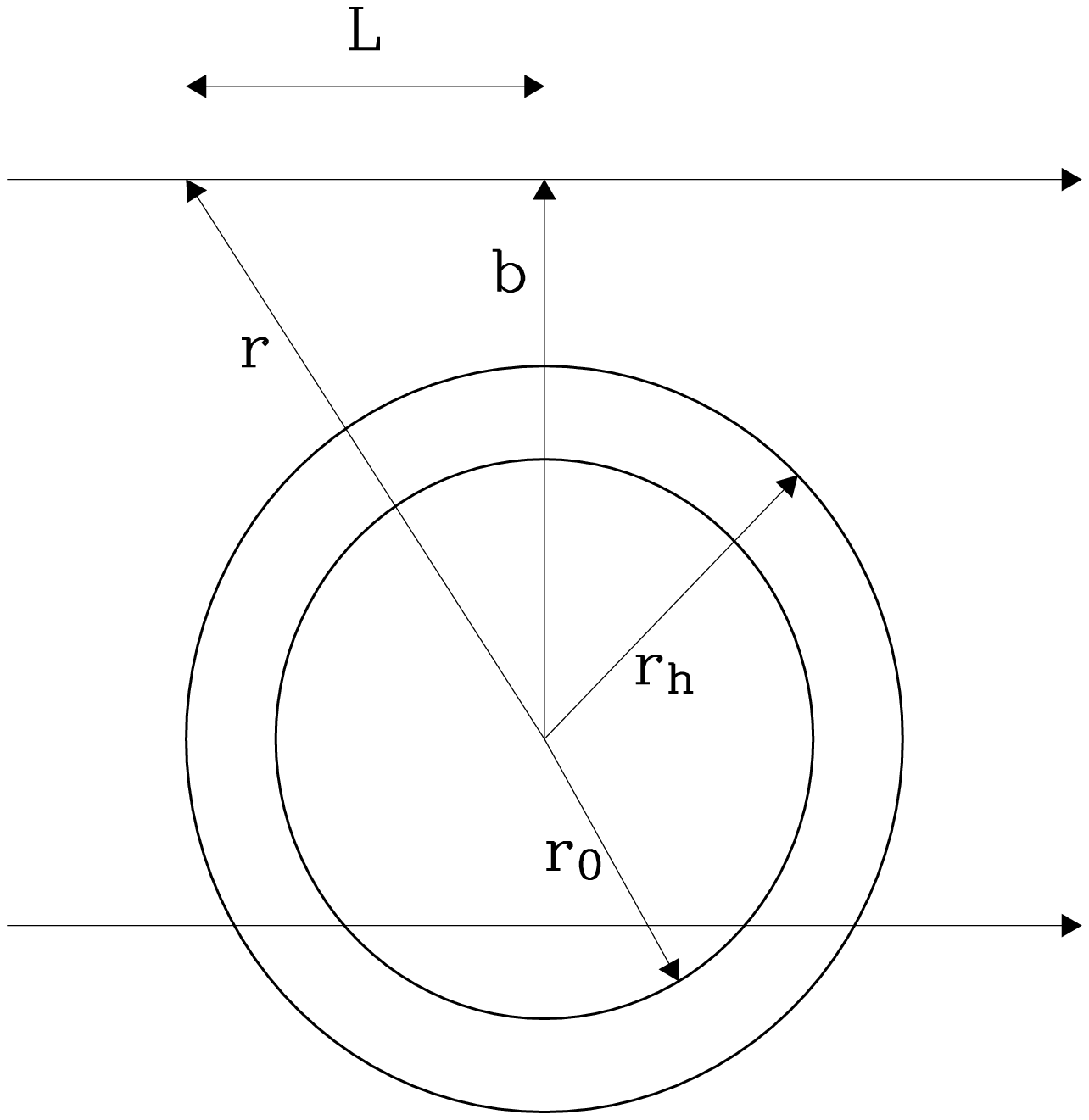,width=4cm,angle=0}
 \caption{The geometry of a light beam passing a Hernquist lens,
  indicating the distance of a photon from the centre of the source,
  $r$, the distance of closest approach, $b$, the radius of the
  Newtonian bubble, $r_0$, and the core radius of the Hernquist
  profile, $r_h$. Top: illustrates situation when $r_h<r_0$. Bottom:
 illustrates situation when $r_h>r_0$.}
 \label{schem}
\end{figure}

\begin{figure}
 \psfig{figure=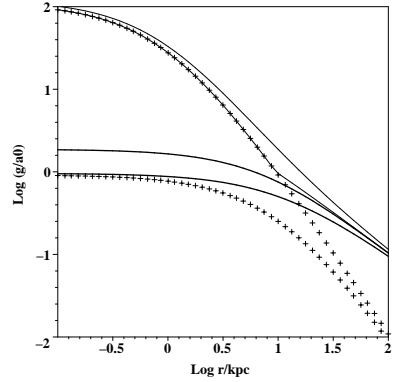,width=5cm} \caption{The rescaled 
 Newtonian field strength $g_N(r)/a_0$ (crosses) and rescaled gravity $g(r)/a_0$ (solid curves) 
 for an unsmooth function $\tilde{\mu}$ (cf. eq.~\protect{\ref{mymu}}) and 
 a smooth function (cf. eq.~\protect{\ref{musmth}}) 
 around Hernquist models of scale length $r_h=1\kpc$ or $11\kpc$ (thin or thick curves) for an elliptical galaxy
 $M=10^{11}\msun$.  The scalar field contribution $\grad \phi c^2/a_0$ is $(g-g_N)/a_0$.
} \label{gpoint}
\end{figure}

Figure \ref{gpoint} compares the gravity (solid curves)
and Newtonian field strength (plus signs) around a galaxy in a compact Hernquist mass model 
($r_h=1\kpc$) and an extended Hernquist model ($r_h=11\kpc$).  For our choice of $\tilde{\mu}$, 
the scalar field contribution starts at small radii if the gravity $g$ is below $a_0$,
it then peaks at a value $a_0/4$ at a radius $r=2r_0-r_h$, and then falls to zero again.

\begin{figure}
 \psfig{figure=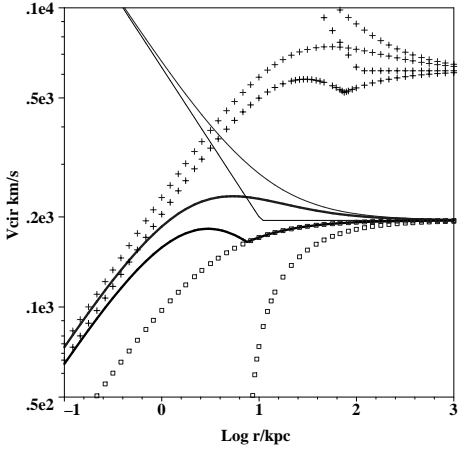,width=6cm}
\caption{ Circular velocity for an elliptical galaxy (lower thick
and thin lines) with $M=10^{11}\msun$ for a point mass model (thin
line) and a Hernquist model (thick line) with scale length
$r_h=3\kpc$ for an unsmooth function $\tilde{\mu}$ (cf. eq.~\protect{\ref{mymu}}) and 
a smooth function (cf. eq.~\protect{\ref{musmth}}); 
The boxes show the contribution of the scalar field in these two choices of $\tilde{\mu}$.  
Likewise for a galaxy cluster with $M=10^{13}\msun$ (crosses)  
for a point mass model and a
Hernquist model with scale length $r_h=30\kpc$. The sharp
break point is the transition radius $r_0$ from strong gravity to
MOND weak gravity.  } \label{vcirpoint}
\end{figure}

Figure \ref{vcirpoint} compares the radial distribution of the circular
velocity for a point-mass and a Hernquist mass distribution in non-relativistic
MOND. In the point mass case we see that 
the velocity curve drops in a Keplerian fashion in the strong, Newtonian limit.
At radii beyond the Newtonian bubble radius, the gravity enters the weak
MOND-limit, and the velocity flattens to $v=v_0$.
The size of the Newtonian bubble scales with $M^{1/2}$, and the terminal
velocity scales with $M^{1/4}$.  The Hernquist models approach the point mass
model at large radii; but at small radii, the circular velocity curves are 
rising instead.  

\subsection{Deflection by the Hernquist model}

We can now find the reduced deflection angle, $\bar\alpha$, for a
light beam passing at a radius $\theta = b/D_L$ from the Hernquist
lens centre. This can again be calculated from equations
(\ref{alpdef}) and ({\ref{gperpdef}).  Substituting in equation
(\ref{herngrav}) we find
 \beq\label{herndefl}
    {\baralpha \over \theta_0 \eta^2} =
\left[{\theta_0 \theta u + \theta_0 \theta_h H_{0u} \over \theta^2-\theta_h^2} + H_{1u} \right],
 \eeq
where
 \beq\label{Hjk}
    H_{jk} =\cases{ { \arcsin (j \sqrt{H}) - \arcsin (k \sqrt{H}) \over \sqrt{H}},
           & $H \equiv 1-{\theta_h^2 \over \theta^2} \ge 0$, \label{cond1} \cr
                    { {\rm arcsinh}(j \sqrt{|H|}) - {\rm arcsinh}(k \sqrt{|H|}) \over \sqrt{|H|} },
           & $H<0$ \label{cond2} \cr}
 \eeq
where
 \bey
    u&=&\sqrt{\left(1-{\theta_h \over \theta_0}\right)^2
    -{\theta^2 \over \theta_0^2}} \label{cond3}, \qquad
    |\theta|<\theta_0-\theta_h,\\
    &=& 0 \qquad {\rm otherwise}. \label{cond4}
 \eey
The different conditions in equations (\ref{cond1}) to (\ref{cond4})
correspond to different ray paths through the Hernquist lens,
illustrated by Figure \ref{schem}. For instance, the upper ray on the
top panel is outside the Hernquist scale length corresponding to
$|\theta|>\theta_h$, ie $H>0$, and is also outside the Newtonian bubble, ie
$|\theta|>\theta_0-\theta_h$. This latter condition also requires
$\theta_h<\theta_0$, ie the Hernquist length is smaller than the
Newtonian bubble.

The lower ray on the top panel illustrates a different regime,
where $|\theta|>\theta_0-\theta_h$; now the ray passes through the
Newtonian bubble, corresponding to the condition given by equation
(\ref{cond3}).

On the other hand, the bottom panel has $\theta_h>\theta_0$, ie
the Hernquist length is larger than the Newtonian bubble. In this
figure therefore, the condition in equation (\ref{cond4}) always
applies. Nevertheless, the two rays obey the conditions $H>0$ and $H<0$ respectively, as
one ray passes through the Hernquist length and one does not.

In summary, the conditions in equation (\ref{Hjk}) 
govern whether a ray passes through a Hernquist
length of the lens; the conditions in equation (\ref{cond3}) and
equation (\ref{cond4}) govern whether a ray passes within a
Newtonian bubble which is larger than a Hernquist length.

In the limit that $\theta_h=0$, we find
  \be
   H_{0u} = H_{1u} - {\pi \over 2} = - \arcsin u, ~
u =\cases{ \sqrt{1-\left(\frac{\theta}{\theta_0}\right)^2},
            & $|\theta| < \theta_0$, \cr
         0, & $|\theta| > \theta_0$,\cr}
 \ee
which recovers the results for the point-mass (cf. eq.~\ref{alphapt}).

\subsection{Amplification by a Hernquist model}

We can find the amplification $A$ for the Hernquist lens from the lens equation
\beq
A^{-1} = {\theta_s d\theta_s \over \theta d\theta} = \left(1-{\baralpha \over \theta}\right)
\left(1-{d\baralpha \over d\theta}\right), 
\eeq
where $\baralpha={D_{ls} \over D_s} \alpha$ is given by eq.~(\ref{herndefl}).
The amplification diverges when ${d\theta_s \over d\theta}=0$
(condition for radial arc) or $\theta_s=\theta - \baralpha=0$ (condition for Einstein
ring).  In the latter case, the critical radius $\theta_c=y_c \theta_0$ is given by
${\baralpha \over \theta_c} =1.$
Substituting in eq.~(\ref{herndefl}), we have
 \beq
  {1 \over 4\xi} = { u_c +  H_{0u} y_h y_c^{-1} \over y_c^2-y_h^2}
    + { H_{1u} \over y_c},
 \eeq
where $y_h \equiv \theta_h / \theta_0$ and
  \bey
    u&=&\sqrt{\left(1-y_h\right)^2-y_c^2} \qquad 1-y_h>|y_c|\\
    &=& 0 \qquad {\rm otherwise},
  \eey
and $H_{jk}$ is a function of $H = 1-\frac{y_h^2}{y_c^2}$ as in eq.~(\ref{Hjk}).
The above $y_c(\xi)$ relation can be used to weigh the lens mass $M$, and hence measure
the $M/L$ of the lens.  Basically
 \beq
    \sqrt{ G M \over a_0} = r_0 = {\theta_c D_l \over y_c(\xi,y_h)},~\xi={D_l D_{ls} a_0 \over D_s c^2},
 \eeq
where the lens distance measure $D_l$, $\xi$ and the critical line radius $\theta_c$
are all observables. 
In the point mass limit we have $y_h=0$, and 
we find $H=1$, $u=\sqrt{{\rm max}(0,1-y_c^2)}$,
$H_{1u}={\pi \over 2}-\arcsin(u)=\arcsin{{\rm max}(1,y_c)}$, so the critical
line satisfies a simpler relation (cf. eq.~\ref{xipt}).
Note the functional dependence of the rescaled critical radius $y_c$ and 
lens distance measure $\xi$ is fairly complex
even with our simplest choice of the free function $\tilde{\mu}$ (cf. eq.~\ref{delta_mu}).  
It is therefore analytically challenging to work with 
more general functions for $\tilde{\mu}$.

Figure~\ref{ycdata} illustrates how the critical radius increases
with the distance parameter $\xi=\eta^2/4$ for various values of the
Hernquist length scale $r_h$; more concentrated models have bigger critical radii,
with the biggest in the point mass limit.  At small lens distances, we find that the critical
radius of an extended Hernquist model can be very small; this is due to the fact that the
Hernquist model has a surface density that diverges only logarithmically at small
radius. For example, for a very extended Hernquist model with $y_h
\gg 1 \gg y_c$, we have $u=H_{0u}=0$, $H \sim -{y_h^2 \over
y_c^2}<0$, $\eta^{-2}=H_{1u}/y_c \sim y_h^{-1} {\rm arcsinh} {y_h
\over y_c}$. Hence $
        {\theta_h \over \theta_c} \rightarrow
        \sinh\left[ \left({G M \over r_h^2 a_0}\right)^{-1/2}
        {D_0 D_s
        \over 4 D_l D_{ls}} \right] \gg 1$;
the critical radius is relatively small for an extended
Hernquist model deep in weak gravity regime where $\left({G M
\over r_h^2 a_0}\right) \ll 1$, and $D_0 = c^2/a_0 \gg D_l$.

The convergence $\kappa$ and shear $\gamma$ can also be derived
analytically; the expressions are somewhat too lengthy to be produced
here.  Instead we illustrate $\kappa$ and the inverse of amplification $1/A$ 
as a function of the impact parameter $b$ for a hypothetical example of a Hernquist-profile galaxy
cluster lens in Figure~(\ref{hernk}).  Comparing with the point lens
case, the convergence is much larger at small impact
parameter for the Hernquist case.  At very large radius, the
Hernquist model approaches the point lens case.
\begin{figure}
 \psfig{figure=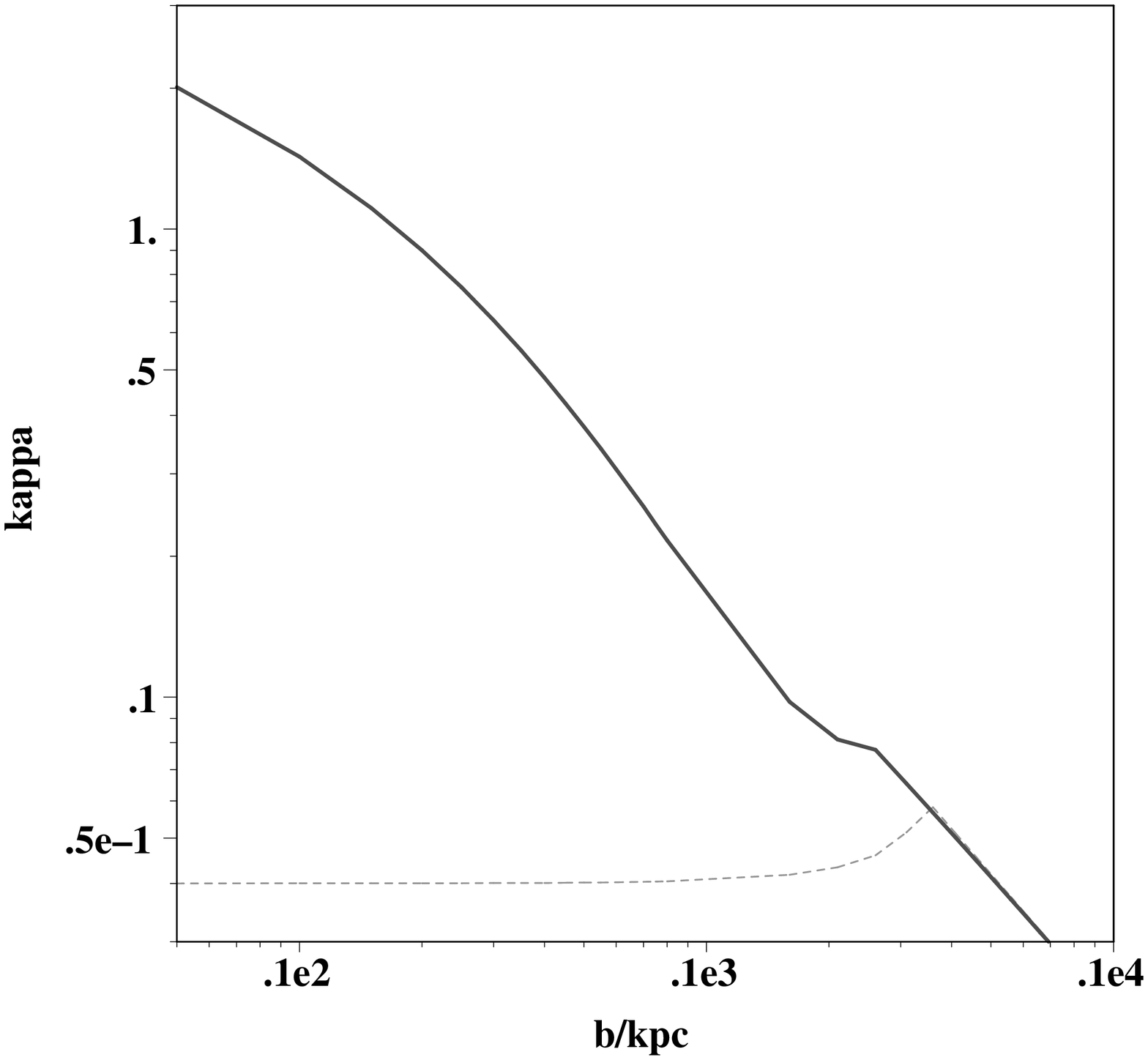,width=4cm}
 \psfig{figure=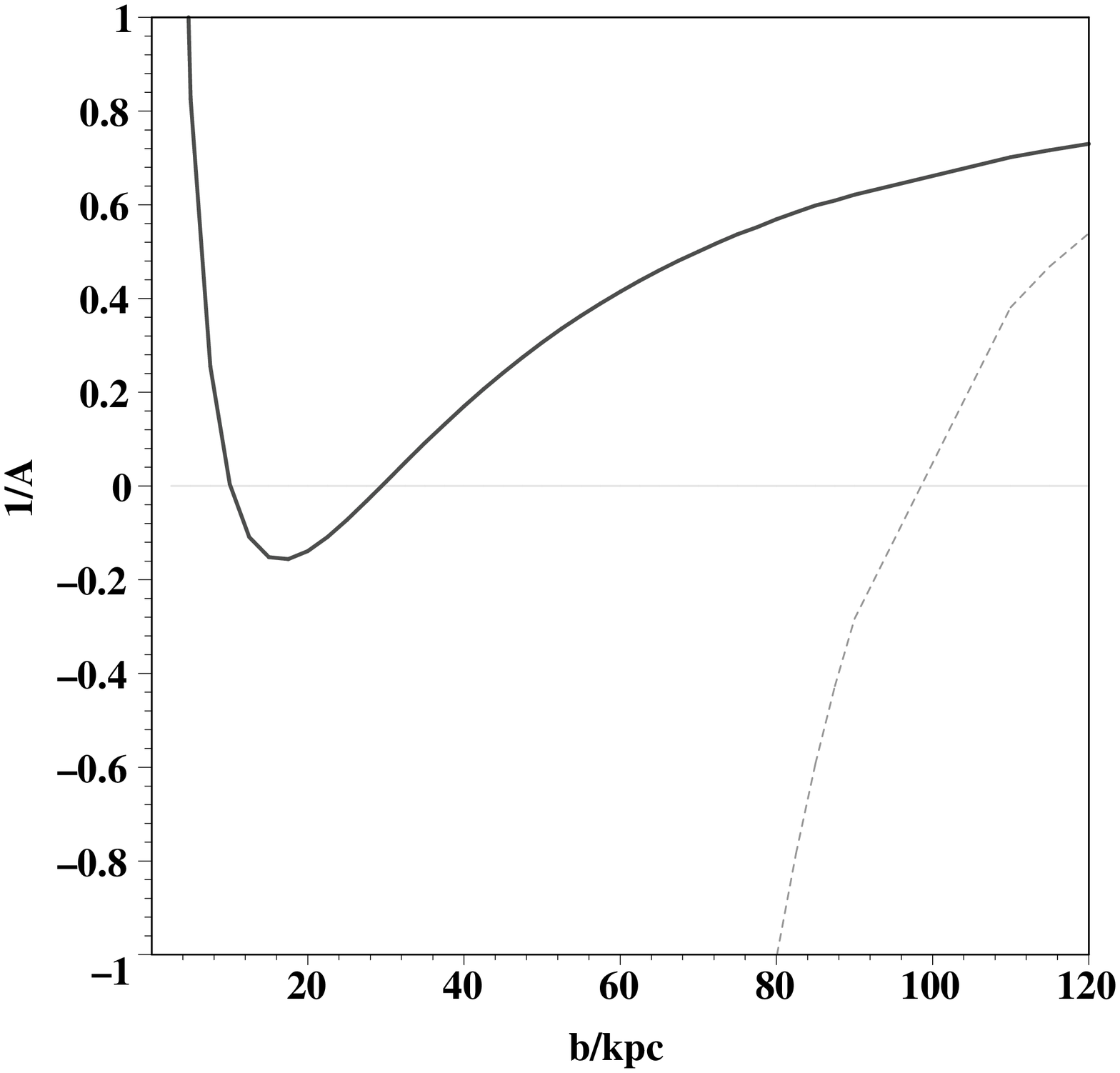,width=4cm}
 \caption{shows two models for a cluster of mass $1\times 10^{14}\msun$ at $D=0.5$Gpc with 
 a Hernquist scale $r_h=100$kpc (solid line) or $r_h=0$ (dashed line). 
 Panel (a): Lens convergence $\kappa$. Panel (b): for the inverse of the amplification
$\left[(1-\kappa)^2-\gamma^2\right]$.  The Hernquist cluster model is only
partly in the weak regime and can create 3 images very close to centre.
A radial arc and a tangential arc can form near the critical points at $b=15$kpc and
35kpc for the Hernquist model. Note that the arcs move inwards
when going from a point model to Hernquist model. 
}
 \label{hernk}\label{hernamp}
\end{figure}

\begin{table*}
\centering
\caption[ ]{Derived mass ratio $M/M_*$ for CASTLES double image lenses.
Lens and source redshifts are listed,
together with angular diameter distance to the lens and the
rescaled lens geometry parameter $\xi$, and lens stellar mass (estimated from evolution and K-corrected I magnitude,
except for two systems from R and K magnitudes as indicated by bracket).  
The lens mass $M$ is calculated using our two techniques (image position
and flux ratio) in TeVeS, for both the point mass model and the
Hernquist model.  For lenses composed of stars of normal stellar populations we expect $M/M_* \simeq 1$.}
\begin{tabular}{l| l l l l l l l l l l}
Lens & $z_l$ & $z_s$ & $D_l$ & ${D_l D_{ls} \over D_s D_0}$ & $M_*$ & $r_h$ & 
${M_{\rm 1,pt} \over M_*}$ &
${M_{\rm 2,pt} \over M_*}$ &
${M_{\rm 1,Hern} \over M_*}$ & 
${M_{\rm 2,Hern} \over M_*} $ \\
CASTLES & & & (Mpc) & $=\xi$ & $(M_\odot)$& (kpc)&
 &  &  &  \\
\hline
Q0142-100 & 0.49 & 2.72 & 1190 & 0.034 & $40.8\times 10^{10}$ & 1.6 &
     0.34 &      0.19 &      0.71 &      0.61 \\
B0218+357 & 0.68 & 0.96 & 1400 & 0.013 & $26.7\times 10^{10}$ & - &
  0.08 &     0.065 &      0.35&      0.32 \\
HE0512-3329 & 0.93 & 1.57 & 1560 & 0.019 & $556.\times 10^{10}$ & - &
0.0025 &       - &     0.04 &       - \\
SDSS0903+5028 & 0.39 & 3.61 & 1010 & 0.032 & $38.0\times 10^{10} (R)$ & - &
      0.7 &       0.53 &       1.10 &       0.95 \\
RXJ0921+4529 & 0.31 & 1.65 & 880 & 0.027 & $3.36\times 10^{10}$ & - &
  50.7 &       41. &       59.7 &       50. \\
FBQ0951+2635 & 0.24 & 1.24 & 730 & 0.022 & $3.09\times 10^{10}$ & 0.32 &
    0.75 &      0.35 &       1.16 &      0.71 \\
BRI0952-0115 & 0.41 & 4.50 & 1070 & 0.035 & $2.68\times 10^{10}$ & 0.29 &
      1.3 &       1.38 &       1.6 &       1.7 \\
Q0957+561 & 0.36 & 1.41 & 980 & 0.027 & $84.4\times 10^{10}$ & 5.23 &
     1.1 &      0.32 &       2.8 &       - \\
LBQS1009-0252 & 0.88 & 2.74 & 1540 & 0.032 & $9.00\times 10^{10}$ & 0.80 &
    1.45 &       1.04 &       2.24 &       1.90 \\
Q1017-207 & 0.78 & 2.55 & 1470 & 0.032 & $7.40\times 10^{10}$ & 1.19 &
     0.53 &      0.42 &       1.19 &       1.49 \\
B1030+071 & 0.60 & 1.54 & 1320 & 0.027 & $16.6\times 10^{10}$ & 1.50 &
   0.45 &      0.23 &       1.45 &       1.8 \\
HE1104-1805 & 0.73 & 2.32 & 1440 & 0.032 & $33.2\times 10^{10}$ & 2.48 &
      2.1 &       1.23 &       3.0 &       1.9 \\
SDSS1155+6346 & 0.18 & 2.89 & 540 & 0.020 & $4.40\times 10^{10} (K)$ & - &
   .43  &       - &      3.8 &       - \\
SBS1520+530 & 0.72 & 1.86 & 1430 & 0.028 & $28.0\times 10^{10}$ & 1.32 &
     0.57 &      0.33 &      0.96 &      0.72 \\
B1600+434 & 0.41 & 1.59 & 1070 & 0.029 & $4.00\times 10^{10}$ & - &
      1.9 &       1.13 &       4.3 &       3.5 \\
PKS1830-211 & 0.89 & 2.51 & 1540 & 0.031 & $14.8\times 10^{10}$ & - &
   0.5 &      0.94 &       1.1 &       1.7 \\
HE2149-2745 & 0.50 & 2.03 & 1200 & 0.032 & $20.0\times 10^{10}$ & 11.4 &
  0.5 &      0.29 &       6.0 &       - \\
SBS0909+523 & 0.83 & 1.38 & 1506 & 0.019 & $655.\times 10^{10}$ & - &
  0.03 &      0.02 &     0.05 &      0.04 \\
\end{tabular}
\end{table*}

\section{Comparing TeVeS predictions to galaxy lens data}

\subsection{Stellar mass of the CASTLES sample}

Now that we have developed models for lensing by point masses and
Hernquist profile lenses, we will apply this formalism to galaxy lens
observations. In particular, we will examine the consistency of the
strong lensing predictions for galaxy lenses in the CASTLES survey
(CfA-Arizona Space Telescope Lens Survey; Munoz et al 1999).

Here we use galaxies with known double/quadruple lensed quasars, with
critical lines estimated from the quasar separation, baryonic mass
inferred from luminosity, and known redshifts from the CASTLES sample
(Kochanek et al, 2000). In the spirit of TeVeS all of the lensing
galaxies have $M/L$ of order unity.  However, to be rigourous, we must
include the K-correction, the luminosity evolution with redshift, and
the possibility of significant gas and extinction from dust.

Kochanek et al (2000, Table 6) have
measured combined K-correction and evolution corrections for CASTLES
lenses, and find a correction for the $I$ band which varies with
redshift ($0.36<z<0.88$ as appropriate for most of our lenses, 
a low density open universe with $\Omega_0=0.3$) 
corresponding to a $\simeq 11$ percent mean offset in
luminosity, which will not impact significantly on our
conclusions about whether TeVeS provides reasonable mass-to-light
ratios.  

To make the interpretation of our lens model more direct, we first
estimate the stellar content of the high-z lens using its observed
I-magnitude and a model for the spectral energy distribution of an old
non-evolving stellar population.  We first estimate the luminosity
without K-correction, or evolution/reddening correction by \beq L =
\left({D_A(z)(1+z)^2 \over 10\pc}\right)^2 10^{0.4 (m_{\lambda_0,
\sun} - m_{\lambda_0, obs})}, \eeq where $D_A(z)$ is the angular
diameter distance to the lens, $(1+z)^4$ gives the dimming effect,
$m_{\lambda_0,\sun}$ is the $\lambda_0$-band magnitude of the Sun.  We
then estimate the stellar mass by the simple formula \beq {M_* \over
\msun} = \gamma\left({\lambda_0 \over 1+z}\right) \times L,\qquad
\gamma\left({\lambda_0 \over 1+z}\right)= \gamma(\lambda_0)
10^{0.4\gamma_1 z}, \eeq where $\gamma\left({\lambda_0 \over
1+z}\right)$ is the estimated mass-to-light ratio of a typical nearby
elliptical galaxy observed at the emitting wavelength ${\lambda_0
\over 1+z}$; e.g., observations at I-band $\lambda_0=8140A$
corresponds to light emitting at $z=0.5$ with a rest-frame wavelength
${\lambda_0 \over 1+0.5}=5000A$, approximately the V-band.  The
parameter $\gamma_1$ parametrizes the countering effects of the
K-correction, and the passive evolution/extinction after starbursts;
these effects tend to cancel each other for most cosmologies and
reasonable formation redshift of ellipticals, and models with very
late/very early formation ellipticals give $\gamma_1>0$ and
$\gamma_1<0$ respectively (see Fig.8 of Kochanek et al. 2000).  For
the I-band we set \beq m_{8140A, \sun} = 4.1^{m}, \qquad \gamma
(8140A) = 4 {\msun \over \lsun}, \qquad \gamma_1 \approx 0.  \eeq This
is calibrated using Fig.32 of Worthey 1994, assuming a 12-Gyr old
solar metallicity stellar population; while the I-band luminosity of a
12-Gyr old red elliptical galaxy is more than its V-band luminosity,
it is actually comparable to the rest-frame V-band luminosity of its
5-Gyr old younger counterpart at $z=0.5$.  For three of our lenses
without I magnitude, we use the nearest available band R, H, V or K
magnitude.  We make similar conversions of mass to light using Worthey
models for these bands, and checked against the predicted redshift
dependence of the lens galaxy colours $V-I$, $R-I$, $I-H$ with those
in the literature (cf. Fig.4 of Keeton et al. 1998, Fig. 1 of Kochanek
et al. 2001).
\footnote{We assume
$\gamma(20500A)=1$ and $\gamma_1 \sim 0.4$ for K band, 
$\gamma(16000A)=0.8$ and $\gamma_1 \sim -0.5$ for H band,  
$\gamma(5550A)=7$ and $\gamma_1 \sim 2$ for V band, and
$\gamma(6500A)=5$ and $\gamma_1 \sim 0$ for R band.
The solar absolute magnitudes are from Allen's Astrophysical Quantities. 
These relations take account of dust in galaxies.
}

Galaxies tend to be more dusty and gas rich at higher redshift.  Falco
et al. (1999) found that a rest-frame differential reddening $\Delta
E(B-V) \sim 0.07^m-0.1^m$ on average for 23 lenses (mostly early-type
galaxies).  McGough et al (2005) have compiled extinction for 25
CASTLES lenses (see their Table 1); the extinction observed at I-band
of most lenses is $\Delta A_I \sim {5550 A \over 8140 A (1+z)^{-1} }
\times 3.1 \Delta E(B-V) \sim 2(1+z)\Delta E(B-V) \le 0.8^m$, i.e.,
less than a factor of two correction of the luminosity.  Note that we
exclude five lenses with $\Delta E(B-V)>0.4/(1+z)$, which are mainly
high redshift late-type lenses.  The consensus seems to be that gas
and dust play insignificant roles in elliptical galaxies in general,
and there is strong evidence against a large dust lane for a lens at
$z=0.84$ (Kochanek et al. 2000b).  In any case, only the far side of a
spherical lens galaxy is affected by the dust lane, hence the dust effect
on the total luminosity is likely {\it mild}, less than a factor of two.

\subsection{Critical lines and lens geometry test}

Now we compare the predicted position of critical radii in TeVeS with
CASTLES data.  Figure \ref{ycdata} shows the predicted scaling of the critical
line opening angle, $y_c = \theta_c/\theta_0$, with the lens
geometry factor, $\xi=\eta^2/4$. It is interesting that the range of
galaxies in this survey all lie in the strong-gravity regime,
where $\xi \ll 1$.  This is due to the fact that
the angular diameter distance in a TeVeS universe has a maximum value
which is much smaller than the $D_0$ scale of TeVeS
(cf. Fig.\ref{fig:DAz}).  Therefore all of the lenses in the sample
have a small $\xi$, and are never in the purely weak gravity regime.
This "coincidence" is actually a natural consequence of the small value for $a_0$.  
has also been anticipated by Sanders (1999), who noted in passing that 
MOND effects become important below a threshold in the surface density 
of a galaxy cluster, while the minimal surface density of 
image-splitting galaxy clusters is always above this MOND threshold.  
Hence the deep-MOND effect is argued to be small.

Secondly we note that the mean position of galaxies in the plot lie on
the point-lens prediction.  This appears to be a success for TeVeS,
where the stellar mass can produce reasonable size Einstein rings.  On
the other hand real galaxies are not point masses.  Furthermore there
are many galaxies which are outliers in this distribution; this
scatter in lens bending power is not expected in a TeVeS universe, as
the displayed TeVeS predicted curve should be universal if galaxies
can be approximated as point lenses.

Thirdly the scatter of observed lenses cannot be explained by extending
the lens via a Hernquist model.  As shown in
Figure~\ref{ycdata}, Hernquist models have lower bending power,
and so cannot explain the images with large separation.

\subsection{Double image lenses}

We now conduct a more detailed study of double-image image lenses from the
CASTLES survey in the context of TeVeS theory with the point and Hernquist
models. Here we examine only lenses with almost co-linear double
images, as non-co-linear images and quadruples
cannot be modelled in detail by our spherical models.

\begin{figure}
\psfig{figure=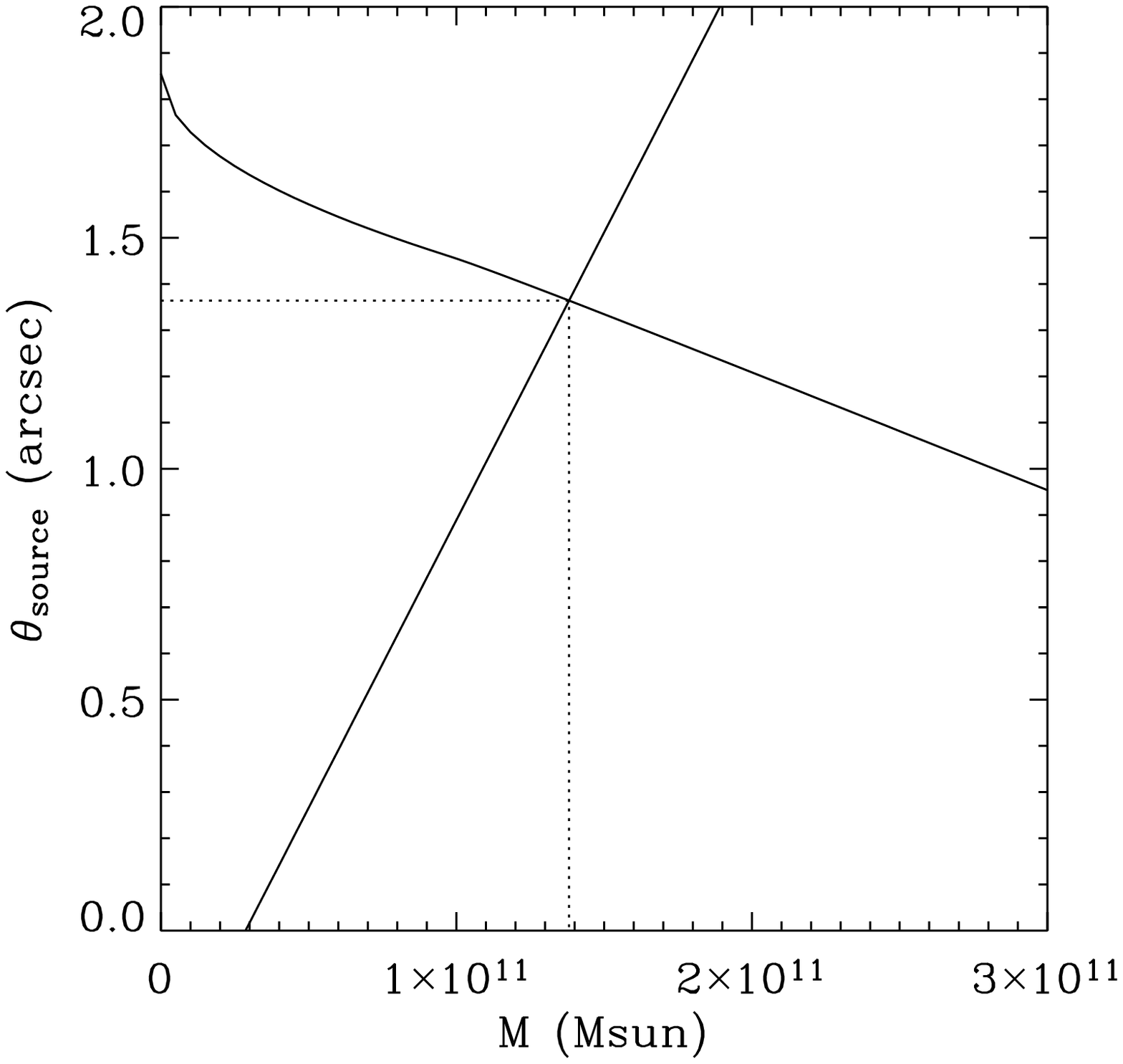,width=4cm,angle=0}
\psfig{figure=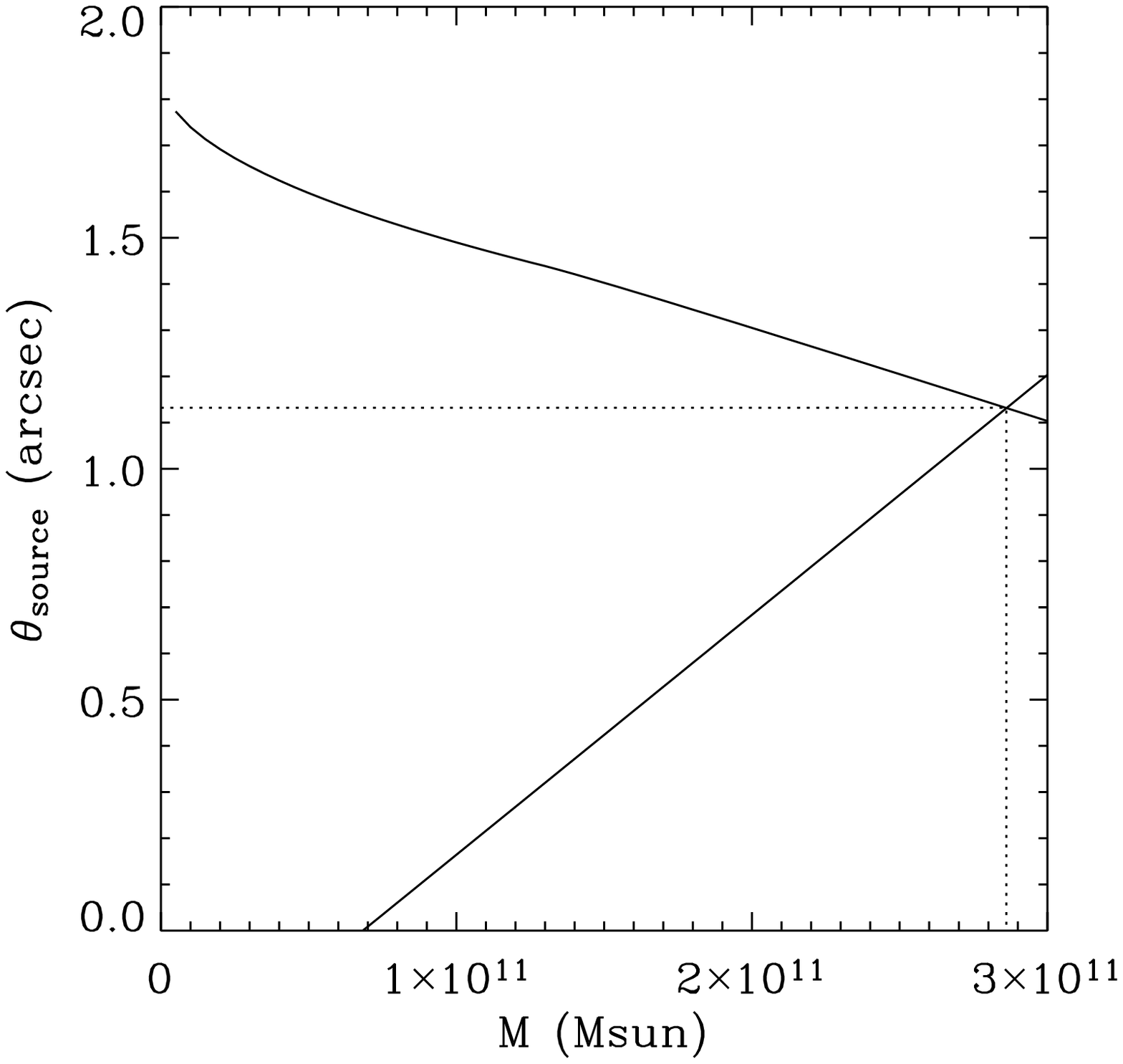,width=4cm,angle=0}
\caption{Upper panel: The TeVeS lens mass for $Q0142-100$ using the
image position technique; here we assume a point mass lens model. The
two curves represent the permitted mass and source position for the
two images; their intersection provides a unique consistent TeVeS mass
for the object.  Lower panel: The TeVeS lens mass for $Q0142-100$
using the image position technique, with Hernquist scale length of
1.6kpc. }
\label{defl}
\end{figure}

\begin{figure}
\psfig{figure=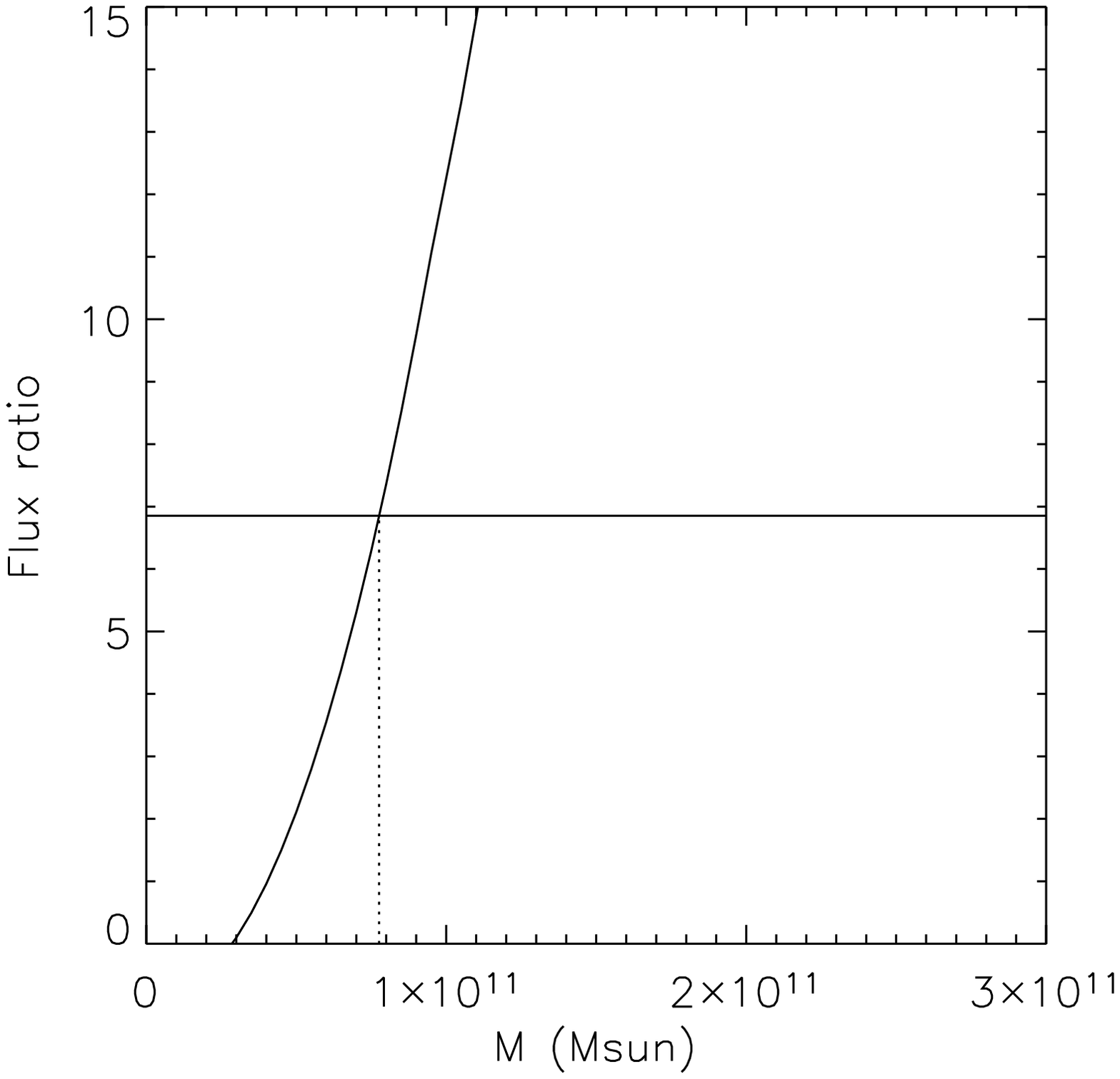,width=4cm,angle=0}
\psfig{figure=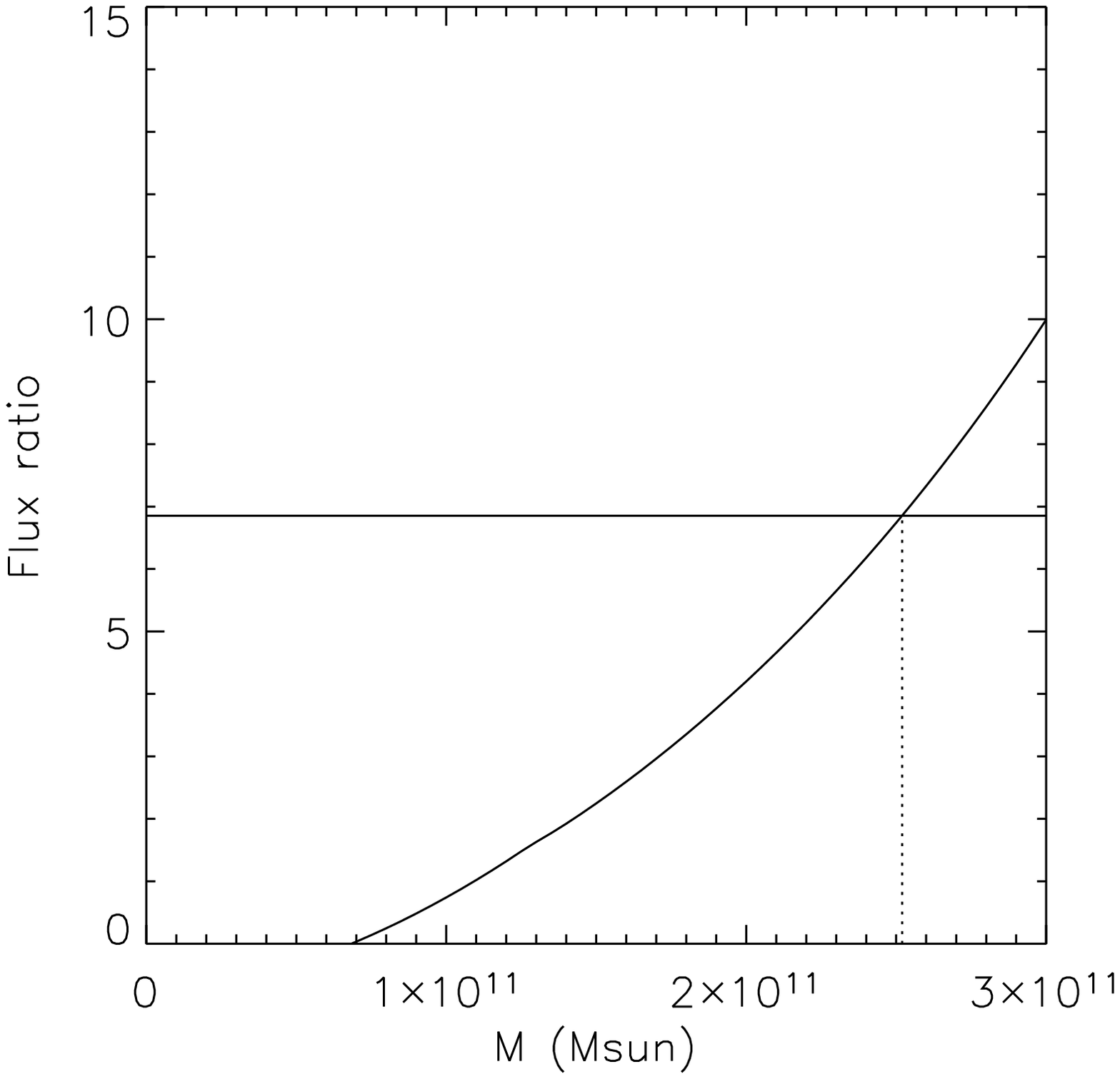,width=4cm,angle=0}
\caption{The TeVeS lens mass for $Q0142-100$ using the flux ratio
technique, with a point mass model. The curve represents the permitted
mass and flux ratio for the two images; the straight line provides the
measured flux ratio. The intersection provides a determination of the
TeVeS mass for the object.  Lower panel: The TeVeS lens mass for
$Q0142-100$ using the flux ratio technique, with Hernquist scale
length of 1.6kpc. }
\label{flux}
\end{figure}

\subsubsection{Source position method}

We determine the lens mass by two methods. The first is by inverse ray
tracing, i.e. from one of the image positions we can predict the
source position as a function of lens mass, using e.g. equation
(\ref{lenseq}). By matching source positions from both images, we
therefore measure the mass according to the TeVeS/MOND theory.

An illustration of this technique is given in Figure \ref{defl} for CASTLE
lens $Q0142-100$. This system includes two images at angular distances
1.9'' and 0.4'' from the lens, with source redshift 2.72 and lens
redshift 0.49. In the figure we see that, for a TeVeS point-mass, we
can calculate predictions for the real, pre-lensed angular position of
the source as a function of mass for each of the two images; where
these predictions intersect we have a unique measurement of the lens
mass, which is here found to be $M=(1.38\pm 0.002)\times10^{11}M_\odot$.

The exercise can be repeated with a finite Hernquist scale-length
$r_h$ for the lens. The bottom panel of Figure \ref{defl}
demonstrates that the resulting mass estimate is quite sensitive
to $r_h$; for a scale length of 1.6kpc for our lens galaxy, the
mass estimate is increased to
$(2.86\pm0.004)\times10^{11}M_\odot$. In order to find $r_h$ in
each case, we use the half light radius for the galaxy, $r_e$ as
given for CASTLES lenses by Kochanek et al (2000) and using the
relation of Hernquist (1991), $r_h=r_e/1.82$ to relate $r_e$ to
$r_h$.

\subsubsection{Flux ratio method}

The second method of measuring lens mass is to predict an
amplification map on the image plane as a function of mass. This
can be achieved by calculating the rate of change of deflection
angle $\alpha$ with the image position $\theta$ as a
function of mass. From this one can calculate the rate of change
of apparent position with respect to the unlensed source position
$\theta_s$, hence the flux ratio of two images is predicted by 
\begin{equation}
{A_1 \over A_2}=\frac{\theta_1}{\theta_2} 
\left( { 1-\frac{\partial \alpha_1 }{ \partial
\theta_1} \frac{ D_{LS} }{ D_S} 
\over 1-\frac{\partial \alpha_2 }{ \partial
\theta_2} \frac{ D_{LS} }{ D_S} } \right)^{-1}.
\end{equation}
Figure \ref{flux} shows this predicted ratio as a
function of mass for the example double-image lens $Q0142-100$, for a
point-mass TeVeS model; also shown is the empirical flux ratio between
the two lenses. We find that the TeVeS mass consistent with the
observed flux ratio is $(7.8\pm0.09)\times 10^{10}M_\odot$. This is
similar in magnitude to the mass found from the source position method
above, but is clearly not consistent with it in detail. In order to
obtain consistency, one requires a Hernquist model with
$r_h=2.3$kpc. In contrast, the value of $r_h$ expected from
half-light measurements is 1.6kpc, so modelling the extended nature of
the lens mass cannot be used to account for all of the difference
between the two estimates. One may see this as a potential difficulty
for TeVeS; alternatively we should note the simplicity of the model we
are using (spherically symmetric, specific profile) which may account
for some or all of the 43\% difference between the two estimates.

The bottom panel of Figure \ref{flux} shows the equivalent test with
the measured $r_h=1.6$kpc; in this case we find a TeVeS mass of
$(2.52\pm0.03)\times 10^{11}M_\odot$. In this case, the mass is closer
in magnitude to the TeVeS mass found via the image position technique
($\simeq13$\% difference), but is not consistent in detail; again, the
simplicity of the lens model may be the cause of this.

\subsubsection{Results for CASTLES lenses}

We have applied the two TeVeS mass estimation techniques to a set
of double image lenses from the CASTLES survey (Table 2), for both
point mass and Hernquist models. The lenses were selected to have
known redshifts for source and lens, known magnitudes in F814, and
to have only two images. We examine the TeVeS mass of these lenses
in relation to their absolute luminosity (from F814 magnitude) in
Figure \ref{fig:ml}, for point lens and Hernquist lens cases with
the position method. Seven of the lens galaxies do not have
published half-light radii; for these we ascribed the median
Hernquist length of the galaxies with known radii, $r_h=1.8$kpc.

We note that there is an apparent correlation between the TeVeS mass
and the luminosity in the F814 band in both the point and Hernquist
cases; this is broadly consistent with TeVeS, which predicts a strong
correlation between mass and luminosity. However, there are some very
significant outliers, leading to a negative Pearson's correlation
coefficient (-0.12 for point-mass; -0.04 for Hernquist).

Finally we can derive the ratio of lens mass vs. stellar mass for each
galaxy, using lens masses from either the position method or flux
ratio method.  In Figure \ref{mlml} we show the mass-to-mass ratios
$M_{\rm lens}/M_*$ from both of these methods, using the Hernquist
model.

Three points of interest arise from this plot. Firstly, we note that
the mass-to-$M_*$ ratios calculated using the two independent methods
closely agree, i.e. $(M/M_*)_{\rm Flux}\simeq(M/M_*)_{\rm position}$. This
shows that TeVeS is working well in terms of being a theory which
appropriately includes gravitational lensing; if there were not a good
level of agreement, it would suggest that different lensing
predictions were not consistent in TeVeS. As it is, TeVeS is obtaining
consistent results from lensing distortions of position and
magnifications.  Nevertheless, for several lenses, e.g.,
SDSS1155+6346, the flux ratio cannot be reproduced with any lens mass,
hence the empty entries in Table 2.  This might reflect the fact that
the flux ratios have been perturbed by microlensing or substructures.
Overall the source position method is more reliable than the flux
ratio method.

The second point concerns the TeVeS mass and absolute luminosity for
both point and Hernquist models.   There is a strong correlation between
the point and Hernquist mass estimations themselves (see Figure
\ref{fig:mm}).  We can see that point lens models underpredict
the required baryonic mass for a given set of image positions in
relation to the Hernquist model, by a mean factor of 3.6. This
confirms the value of extending our analysis to the Hernquist model.

The third point of interest from this plot concerns the range of
mass-to-$M_*$ ratios measured. All but two of the lenses are found to
have $M/M_*$ between 0.5 and 2; this is a reasonably concentrated
distribution.  However, we note the existence of extreme outliers such
as RXJ0921+4529 with $M/L=239$ and HE0512-3329 with $M/L=0.16$
(cf. Table 2 for the mass ratio $M/M_*$).

\begin{figure}
\psfig{figure=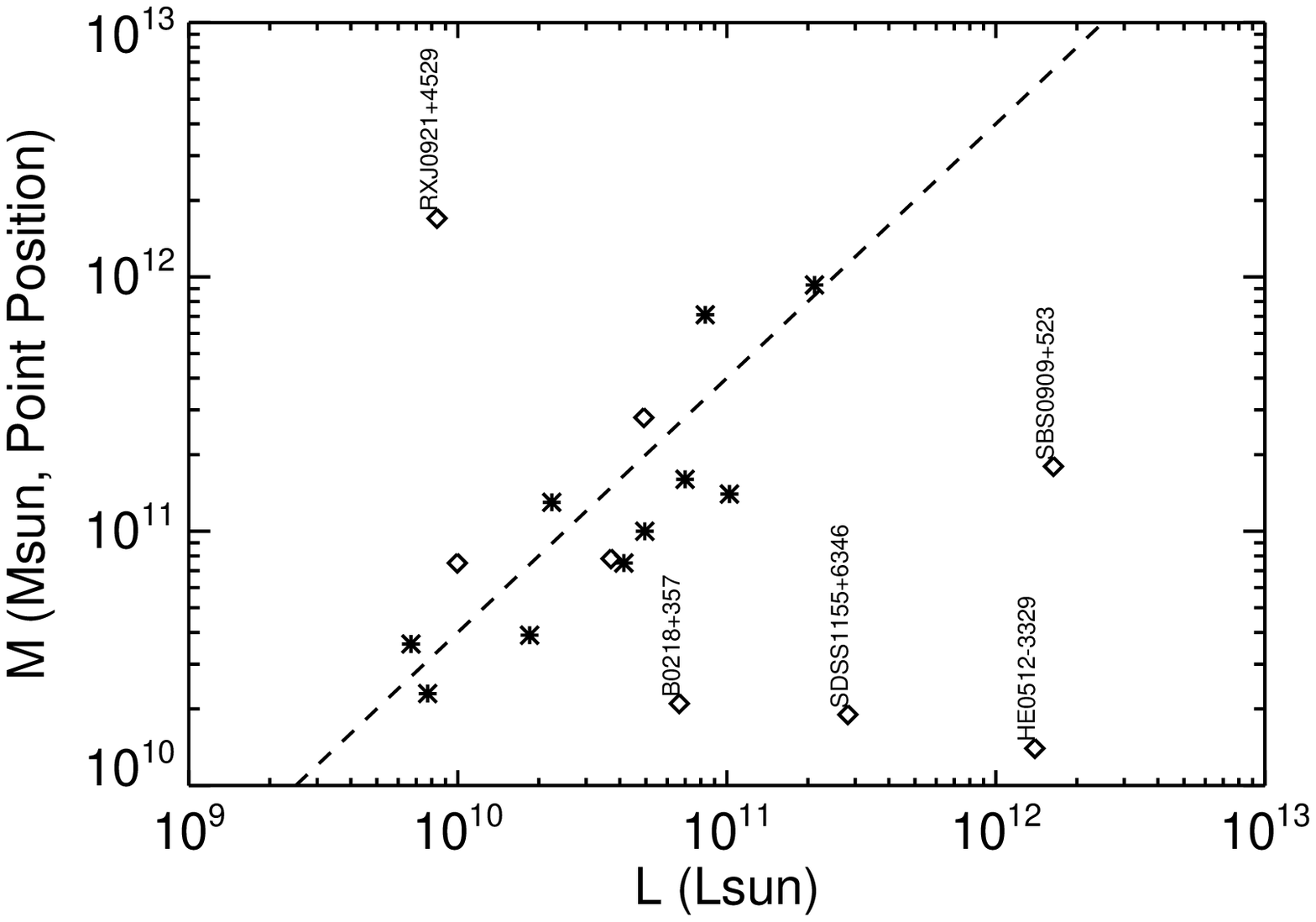,width=7cm,angle=0}
\psfig{figure=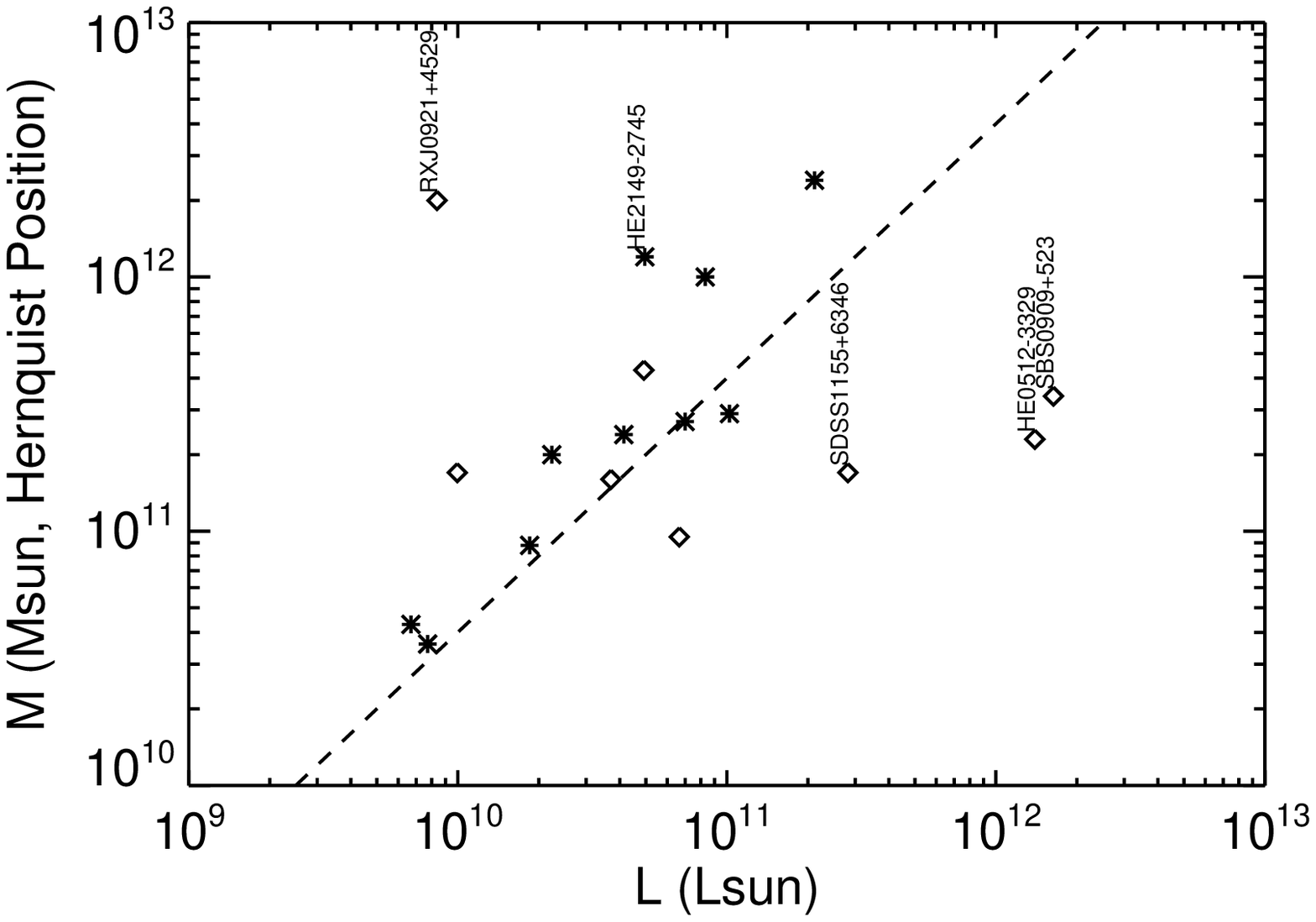,width=7cm,angle=0}
\caption{Upper panel: TeVeS lens mass for double image CASTLES lenses
using the image position technique (point mass), as a function of (converted) F814
absolute luminosity for these lenses. The lines indicate 
$M/L=4$ (dashed).  Lower panel: TeVeS lens mass for double image CASTLES lenses
using the image position technique (Hernquist), as a function of F814
absolute luminosity for these lenses. Open circles show lenses where
the median Hernquist length has been used; for easy comparison, these
lenses are also displayed with open circles in the upper panel, where
Hernquist length is irrelevant.  Note five outliers indicated by 
five open diamonds.}
\label{fig:ml}
\end{figure}

\begin{figure}
\psfig{figure=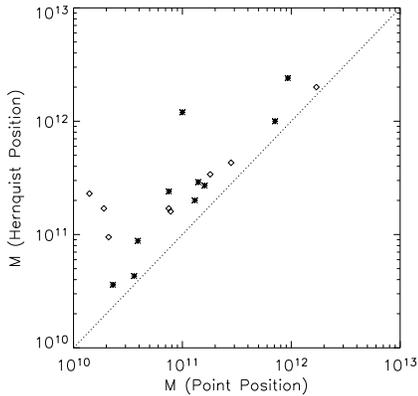,width=6cm,angle=0} \caption{Comparison of
TeVeS masses obtained using the image position technique, for point
masses and Hernquist models. The dotted line indicates equality of
mass; open circles show lenses where the median Hernquist length has
been used.}
\label{fig:mm}
\end{figure}

\begin{figure}
\psfig{figure=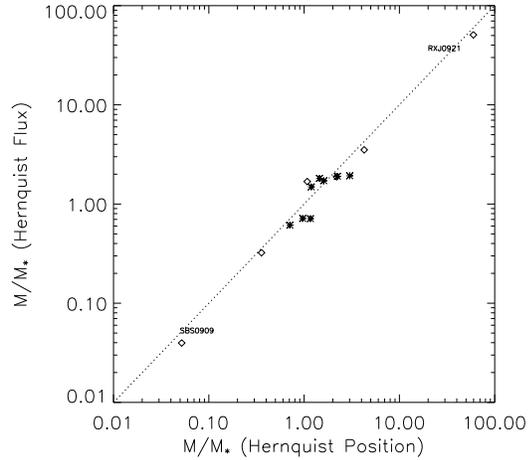,width=7cm,angle=0} \caption{The lens mass to stellar mass ratio 
$M/M_*$ from two methods for Hernquist profiles using realistic length scales. The
dotted line indicates equal mass-to-light ratios; open circles show
lenses where the median Hernquist length has been used.  The data are 
expected to be tightly clustered around $M/M_* \sim 1$ in MOND/TeVeS.}
\label{mlml}
\end{figure}

\section{Discussion}

\subsection{RXJ0921+4529}

As noted in the last section, there are several lenses with anomalous
$M/L$.  In particular the galaxy RXJ0921 with luminosity
$L=0.84\times10^{10}L_\odot$ requires either a point mass of
$1.7\times10^{12}M_\odot$ or a Hernquist profile mass of $2.0\times
10^{12}M_\odot$.  This lens is therefore required to have a $M/L \sim
240$, which could seriously challenge the TeVeS theory.

Nevertheless, several explanations are initially available in defence
of TeVeS for this system. The first possibility is that this lens is
severely dimmed by obscuration of dust; however, the $I-H$ colour of
the lens appears normal, which would not be expected in the presence
of dust.

Mu\'noz et al. (2001) also note an extended emission source B' offset
from the B image for this system, and there is no extended counterpart
in the image A.  An important although unlikely explanation is that A
and B are a pair of independent quasars with small separations and the
same redshift, and the quasar B is off-centred from its host galaxy B'
as well.  A more plausible explanation is that B' is a member galaxy
in the same cluster, projected by chance near one of the two split
images of a lensed quasar.

Another suspicion would be that this outlying mass is due to the
use of too large a Hernquist length (as this is a lens where the
median $r_h$ is used); however, the fact that the large mass is
found even in the point-mass case allays this concern.

Yet another possibility for the high M/L is the fact that the lens
sits in the middle of an X-ray cluster; this will increase the lensing
distortion associated with the system.  This issue clearly requires
follow-up using lens models more sophisticated than those considered
in this paper, which would be able to model multi-component lenses
with different length-scales.

Another system with a mildly high $M/L$ is B1600+434; however, unlike our
other lenses this is found to be an edge-on spiral galaxy. A Hernquist
model will not provide a good fit to the baryonic matter in a spiral,
so it is unsurprising that this $M/L$ is an outlier. Indeed, with a
point mass model, the mass-to-light is more reasonable (4.5 or 7.5).

In summary, at least one outlier in our study succeeds in overcoming
some of the defences available to TeVeS; however, the presence of a
lensing cluster does call into question the level of disagreement between
TeVeS predictions and data on this lens.

\subsection{Effects of different choices of $\tilde{\mu}$ and $a_0$}

One concern regarding the above result is that we used an unsmooth function $\tilde{\mu}$ (cf. eq.~\ref{mujump}), which
reduces the contribution of the scalar field to exactly nothing in the strong gravity regime. 
Since most lenses have impact parameters inside the Newtonian bubble, letting the scalar field contribute
in the strong gravity could help towards the $M/L$ discrepancy.  
Common choices of $\tilde{\mu}$ include
\beq
\tilde{\mu}  = \cases{  {g / (g+a_0)} \cr
                        {g / \sqrt{g^2+a_0^2} } \cr
                        1 - \exp(-g/a_0). \cr} 
\eeq
Another concern is that $a_0$ could be slightly (by a factor of two in some systems) 
higher than $1.2 \times 10^{-8}{\rm cm}{\rm s}^{-2}$ 
given reasonable errors on disk galaxy rotation curve (RC) data (Sanders \& McGaugh 2002).

To address these concerns we consider a new smooth function $\tilde{\mu}$ so that
\beq
\delta_\phi = {\grad \phi c^2 \over a_0} = {\tilde{\mu} \over 1-\tilde{\mu}}
\eeq
is a monotonic increasing function of $\tilde{\mu}$, closely matching that chosen by Bekenstein
(cf. eq.~\ref{bekenmu}}, as illustrated in Fig.~\ref{fig:mutilde}).

\begin{figure}
\psfig{figure=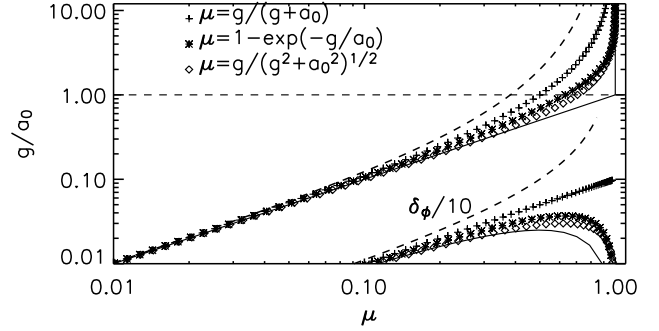,width=9cm,angle=0} \caption{The MOND/TeVeS functions
$\tilde{\mu}$ and $\delta_\phi$ (shifted down by a factor of 10).  
Common choices of $\tilde{\mu}$ are shown in symbols, which
are bracketed by our choices (thin and thick solid curves).  All these choices
converge asymptotically so that $\tilde{\mu}\rightarrow \delta_\phi \rightarrow g/a_0$
in the weak gravity regime (below the horizontal dashed line).}
\label{mus}
\end{figure}

With a bit of algebra it can be shown that
\beq
g = g_N + \grad \phi c^2 = g_N + \sqrt{g_N a_0},
\eeq
and
\beq\label{musmth}
\tilde{\mu} = {\delta_\phi \over 1+ \delta_\phi} 
= {a_0 \over 4 g} \left[ \sqrt{1+{4 g \over a_0}}-1\right]^2.
\eeq
The gravity and rotation curve of a Hernquist galaxy or galaxy cluster 
in such a model are shown in Fig.~\ref{gpoint}
and ~\ref{vcirpoint}.  The curves are much smoother than those in our principal model.  
Here the scalar field contributes even inside the Newtonian bubble (strong gravity regime),
unlike that in our principal model.  Also this model and our earlier choice of $\tilde{\mu}$ bracket
all common choices of $\tilde{\mu}$ (cf. Fig.~\ref{mus}).  

Interestingly, our new choice of $\tilde{\mu}$ also allows 
the deflection of Hernquist lens model to be calculated analytically, and we find 
that the rescaled critical radius $y_c$ satisfies 
 \beq\label{hernsmth}
 {y_c \over 4 \xi } = \left[{ y_c - y_h H_{10} \over y_c^2-y_h^2} + H_{10} \right],
~H_{10} = {\arcsin \sqrt{1-{y_h^2 \over y_c^2}} \over \sqrt{1-{y_h^2 \over y_c^2}}}, 
 \eeq
where 
\beq\label{a0fit}
\xi = {a_0 \over c^2}{D_lD_{ls} \over D_s}, ~ 
y_c = D_l \theta_c \sqrt{a_0 \over G L} (M/L)^{-1/2}.
\eeq
This allows us to estimate the lens mass $M$ by fitting the Einstein ring size; the solution has
to be found by iteration because eq.~\ref{a0fit} is an implicit non-linear function of $M/L$ through $y_c$ and $y_h$.

\begin{table}
\caption[ ]{CASTLES lenses with anomalously large/small ${M_{\rm lens} \over M_*}$.}
\begin{tabular}{l| l}
Lens & Comments \\
\hline
RXJ0921+4529 & 2-image.  Resides in cluster \\
SDS1004+4122 & 2-image.  Resides in cluster \\
B1600+434    & 2-image.  Edge-on spiral lens with dust lane\\
MG2016+112   & 3-image.  One or two lens planes \\
B2045+265    & 4-image.  Source near cusp caustic \\
\hline
HE0512-3329   & 2-image.  Gas-rich lens. \\
SBS0909+532   & 2-image.  Early-type, little gas/dust \\
B0218+357     & 2-image.  Lens is likely a spiral galaxy.\\
RXJ1131-1231  & 4-image.  V-magnitude. Cusp caustic.  Elliptical.\\
B1933+503     & 10-image components.  Fits $R^{1/4}$-law. \\
\end{tabular}
\end{table}

\subsection{Constraints on $\tilde{\mu}$ from a larger sample}

\begin{figure}
\psfig{figure=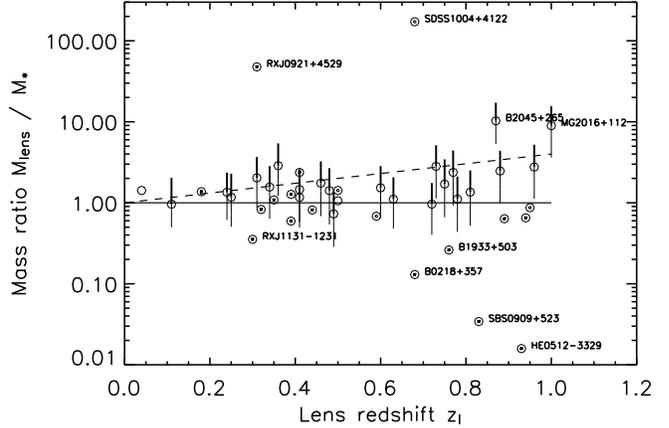,width=9cm,angle=0} \caption{
Shows the lens redshift-dependence of the mass ratio $M_{\rm lens}/M_*$ derived from 
fitting the Einstein ring size of the CASTLES sample (circles).
Circles with error bars are for well-behaved subsample of CASTLES lenses 
which have the Hernquist profile and little dust/gas.
The upper thick part of an error bar brackets the range of uncertainty of 
the modification function $\tilde{\mu}$ 
(as shown in Fig.~\protect{\ref{mus}}) for fixed 
$a_0=a_0^{RC}=1.2\times 10^{-8}{\rm cm}{\rm s}^{-2}$.  
The lower part of an error bar with alternating thickness 
adopts $a_0 = (1-2)a_0^{RC}, (2-4)a_0^{RC}, (4-8)a_0^{RC}$
but fixes $\tilde{\mu}$ as in eq.~\protect{\ref{musmth}}.
Circles with central asterisks are lenses with unresolved scale-length (adopt $1.82\theta_h=0.1\arcsec$).
Circles with smaller circles are very faint lenses with unknown magnitude and scale-length 
(adopt $I=21^m$ and $1.82\theta_h=0.1\arcsec$). 
The solid horizontal line is the expected K-corrected value (unity) for plausible star formation
and passive evolution of ellipticals. 
The dashed line shows the expected K-corrected value but assuming galaxies do not evolve.
}
\label{mlzplot}
\end{figure}

Fig.~\ref{mlzplot} shows the application to a sample similar to those
shown in Fig.~\ref{ycdata}, including all CASTLES double/four-imaged
lenses with measured lens and source redshifts.  Note the very large
scatter overall.  There are galaxies well above and well below the
expected stellar mass, which cannot be simply explained by extinction,
evolution or K-correction.  Interestingly, SDSS1004+4122 is also in a
galaxy cluster (from which the lens redshift was assigned), similar to
the situation of RXJ0921+4529.  At the opposite end, the lens galaxy
of HE0512-3329 is a gas-rich damped Lyman absorber galaxy, barely
resolved at $z=0.93$ with a size $\sim 1\kpc$ comparable to the
resolution of the Hubble Space Telescope (Gregg et al. 2000).
Nevertheless, the stellar mass alone seems to exceed what is required
for lensing.  A similar situation pertains to SBS0909+532, an early
type lens without much gas and extinction (Lubin et al. 2000, McGough
et al. 2005).  Some features of these outliers are compiled in Table
3.

Selecting only a well-behaved subsample (e.g., excluding disky and/or
dusty galaxies, galaxies with unknown half-light radius and/or unknown
magnitude), we find that the scatter is still significant, with
especially large deviations for three high redshift lenses.  The
$M_{\rm lens}/M_*$ shows mild dependence on the lens redshift $z_l$,
suggesting the need for K-correction.  To make $M_{\rm lens}/M_* \sim
1$ for the $z=0.9-1.0$ lenses would require a model where
$\gamma(\lambda_z)$ increases by one decade (i.e., $\gamma_1=2.5^m$)
per unit redshift; this is too steep even for purely K-corrected
models where galaxies do not evolve (dashed line, taken from Fig.8 of
Kochanek et al. 2000).  While it has been suggested before that
MG2016+112 might involve two lenses at two redshifts (Nair \& Garrett
1997), a more recent model found that a single lens plane is
sufficient (Koopmans et al. 2002).  The early type (Sa) lens B2045+265
from the CLASS survey seems to be a robust case of a single lens with
accurate NICMOS near IR data and radio data; the source is near a cusp
caustic, and is split into three images on one side and one image
$2\arcsec$ away on the other side of the resolved lens (Fassnacht et
al. 1999).  The peculiar flux ratios in both radio and near IR among
the three images near the critical curve have been used to highlight
the existence of dark substructures (Keeton et al. 2003).

Another way of describing the residue between mass and light is by
using the theoretically uncertain modification function $\tilde{\mu}$.
We slide the function $\tilde{\mu}$ in between two expressions, say $\tilde{\mu}_1$ and $\tilde{\mu}_2$, which are
that of equation~(\ref{mymu}) and that of equation~(\ref{musmth}) respectively 
by making the linear combination
\beq\label{mix}
{1 \over \tilde{\mu} }= {f \over \tilde{\mu}_1} + {1-f \over \tilde{\mu}_2}, \qquad -\infty<f<1.
\eeq
The results with a slidable $0<f<1$ are shown as the upper thick part of the error bars in Fig.~\ref{mlzplot}.  
Still we find $M_{\rm lens}/M_* \ne 1$ for many systems.  Model with $f<0$ seems necessary.

Yet another possibility is that the value for $a_0=a_0^{RC} = 1.2 \times 10^{-8}{\rm cm}{\rm s}^{-2}$ 
from disk galaxy rotation curve (RC) measurements could be an underestimate.    
Larger values have been suggested in 
galaxy cluster studies (e.g., Pointecouteau \& Silk 2005 and Sanders \& McGaugh 2002).
So suppose the theoretical value for 
\beq
a_0 =(1,2,4,8) \times a_{0}^{RC} =(1.2, 2.4, 4.8, 9.6)\times 10^{-8}{\rm cm}{\rm s}^{-2}.
\eeq
Keeping $\tilde{\mu}$ as in eq.~(\ref{musmth}), adopting bigger $a_0$ has 
the effect of lowering $M_{\rm lens}/M_*$ (shown by the lower segments of the error bars).  
Nevertheless, a good match to $M_{\rm lens}/M_* \sim 1$ still seems difficult. 
A few systems seem to suggest an $a_0$ more than factor of 10 higher than
the standard value $a_0^{RC}$ from rotation curves of disk galaxies.  
In short, the recurring discrepancy between some lenses and MOND/TeVeS predictions 
seems to go beyond the simplest lens models, and remains with many
plausible functions of $\tilde{\mu}$ and plausible values for $a_0$.

A related point is that not all popular MOND $\tilde{\mu}$ correspond to a plausible
Lagrangian for the scalar field.  Zhao \& Famaey (2005) noted that the standard function 
$\tilde{\mu}(g/a_0)=g/\sqrt{a_0^2+g^2}$ makes the gravity $g$ (and all its functions) multi-valued
for the same strength of the scalar field.  This happens for all sharply increasing 
$\tilde{\mu}$, including our principal one (see Fig.~\ref{mus}).  These models are perhaps 
undesirable since the scalar Lagrangian cannot be expressed uniquely by the scalar field strength.
The lensing predictions of TeVeS are likely in a range narrower than depicted by the "error bars" 
in Fig.~(\ref{mlzplot}).

\section{Conclusion}

We have explored key properties of gravitational lensing as it occurs
in Bekenstein's theory TeVeS, a fully covariant modified
gravity theory. We have shown that TeVeS is frequently successful in
predicting gravitational lensing phenomena; however, there are lens
systems where TeVeS appears to fail radically.

We began by giving an account of TeVeS theory.  
Lensing is still governed in TeVeS by a gravitational potential $\Phi$
which enters the metric in a similar way to General
Relativity. However, the interpretation of the gravitational potential
is different; rather than this being due to baryonic and dark matter,
it is solely due to baryonic matter, which generates a gravitational
potential in excess of the GR prediction.  
The usual relations among lensing deflection angle, the lens
equation, convergence, shear and magnification still hold except that 
their values as functions of the impact parameter have changed, as 
we have shown analytically for point-mass lens.  E.g., 
the convergence of a point-mass lens is non-zero in TeVeS.
Analytical lens models are also constructed for 
a Hernquist profile, which should be a good model for baryons 
in most elliptical galaxies.  Conditions for generating arcs
are discussed appertain to galaxy cluster scales.

We then further tested the theory using galaxy lens data from the
CASTLES survey. We find that the observed relationship between
critical lines and lens geometry (cf. Fig.~\ref{ycdata}) would be
largely consistent with TeVeS if lenses were treated as point-lenses,
but there are a handful of outliers.  We also calculated TeVeS masses for
CASTLES double-image lens galaxies, assuming point mass lenses or
Hernquist lenses, and using image positions or fluxes to calculate
mass. We found that TeVeS/MOND provides an acceptable explanation for the
lensing data in general.  But in a handful of cases we obtain too
large or too small mass-to-light ratios (cf. Fig.~\ref{mlml},
Fig.~\ref{mlzplot}, and Tables 2-3).  Another way to put this is that we
observe some cases of deviations from the expected baryonic content;
or the $a_0$ derived from the critical lines of some individual
galaxies deviates markedly from the TeVeS/MOND expectation
(cf. Fig.~\ref{mlzplot}).  More detailed photometric and spectroscopic
data of these outlier lenses (see Table 3) are needed urgently to
clarify the degree of the problem.

On cosmology in the TeVeS context we noted the likely need for  
a cosmological constant (allowed in both TeVeS and GR).
If $\Omega_\Lambda \sim 0.46$ 
it is possible to construct a low density baryonic universe 
that agrees with current cosmological constraints from supernovae and
which equivalently provides an acceptable angular diameter distances 
at low redshifts.  Such an open cosmology might not be the best for TeVeS 
since the {\it sound horizon size at the last scattering is underpredicted}. 
(cf. Fig.~\ref{fig:SN}).

This work has made initial tests of TeVeS using gravitational
lensing. Our analysis of TeVeS in the weak field regime suggests
that our results might be valid for any relativistic theory that
asymptotes to MOND in the weak field regime. Further refinement of
these tests (Shan et al. 2006) will involve examining non-circular TeVeS lenses 
to ascertain whether they fit the double and quadruple imaged lenses
in more detail.  Observed time delays between images of a dozen
systems could also be constraining (Zhao \& Qin 2005). 
Further work will also extend
the analysis to a wider range of density profiles, including the
beta-profile most suitable for analysing cluster lensing. In the
meantime, we see that TeVeS succeeds in providing an alternative
to General Relativity in some lensing contexts; however, it faces
significant challenges when confronted with particular galaxy lens
systems. Finally, we see that gravitational
lensing can be used as a useful approach to distinguish between
theories of gravity, and to probe the functional form of the 
modification function $\tilde{\mu}$.  In the mixed model (cf. Eq.~\ref{mix}) 
the gravity $g$ rises more steeply with $\tilde{\mu}$ for models 
with smaller $f$ and even negative $f$ (cf. Figure~\ref{mus}).  
E.g., the model with $f=0$ (Bekenstein 2005) is more effective in bending light
for the same baryonic mass than models with 
$\tilde{\mu}=x/(1+x)$, $\tilde{\mu}=x/\sqrt{1+x^2}$, 
$\tilde{\mu}= \min(1,x)$ (cf. Figure~\ref{mlzplot}).  
The model with 
$f=0.5$ mimics the simple function $\tilde{\mu}=x/(1+x)$ in the intermediate regime.
Any model with $f>0.5$ is only only inefficient 
in lensing, but also always shows an undesirable peak in the scalar field strength $\delta_\phi^2$.
This leads to unphysical external field effect according to Zhao \& Famaey (2005).
They also find that models with $f \le 0$ have a wide transition zone, which do not fit 
the sharply rising rotation curves seen in galaxies.  Putting all these constraints 
together, it seems that the {\it MOND/TeVeS models with $0 \le f\le 0.5$ have the best chance 
to survive}.  

\section*{Acknowledgements}
HSZ, DJB and ANT thank the PPARC for Advanced Research Fellowships. We
thank Xuelei Chen and John Peacock for helpful comments on MOND cosmology, 
and Chuck Keeton and Chris Kochanek for comments on CASTLES data. 
We also thank Jacob Bekenstein, Benoit Famaey, 
Mario Livio, Stacy McGaugh and Moti Milgram for discussions on MOND, and
the referee Bob Sanders for insightful comments.  
HSZ acknowledges travel and accomodation support from Beijing
Observatory.


\label{lastpage}

\vfill\eject


\begin{thebibliography}{}

\bibitem{BarSch} Bartelmann M., Schneider P., Phys. Rep. 340, 291.

\bibitem{Bekenstein} Bekenstein J. 2004, Phys. Review D., 70, 3509

\bibitem{BM84} Bekenstein J. \& Milgrom M. 1984, ApJ, 286, 7,

\bibitem{BT} Binney J. \& Tremaine S., 1987, Galaxy dynamics, Princeton University Press.


\bibitem[]{} Chiu M., et al. 2005, ApJ, in press, astro-ph/0507332

\bibitem{CiottiBinney} Ciotti L., Binney, J. 2004, MNRAS, 351, 285

\bibitem{} Fassnacht C.D. et al. 1999, AJ, 117, 658 

\bibitem{} Falco E.E. et al. 1999, ApJ, 53, 617

\bibitem{} Famaey B. \& Binney J. 2005, MNRAS, in press 

\bibitem{} Gregg M.D. et al. 2000, ApJ, 119, 2535

\bibitem{Hernquist} Hernquist L., 1990, ApJ, 356, 359

\bibitem[]{} Hao J. \& Akhoury R. 2005, astro-ph/0504130

\bibitem{} Keeton C., Gaudi S., Petters A.O., 2003, ApJ, 598, 138

\bibitem{Kochanek} Kochanek C. et al, 2000, ApJ, 535, 692

\bibitem{Kochanek} Kochanek C. et al, 2000, ApJ, 543, 131

\bibitem{Kochanek} Kochanek C., 2003, ApJ, 583, 49

\bibitem{} Koopmans L. et al. 2002, MNRAS, 334, 39

\bibitem{} Lubin L. et al. 2000, AJ, 119, 451

\bibitem{} McGough C., Clayton G.C., Gordon K.D., Wolff M.J. 2005, ApJ, 624, 118 

\bibitem{Milgrom83} Milgrom M. 1983, ApJ, 270, 365

\bibitem{Milgromexact} Milgrom M., 1986, ApJ, 302, 617

\bibitem{MilgromSanders} Milgrom M. , Sanders R. H. 2003, ApJ, 599, L25,

\bibitem{Munoz} Munoz J.A., Kochanek C., Falco E.E., 1999, Astrophys.
Space Sci. 263, 51.

\bibitem{} Mortlock D., Turner, E.L., 2001, MNRAS, 327, 557

\bibitem{} Nair S. \& Garrett M. 1997, MNRAS, 284, 58


\bibitem{percival} Percival W., et al., 2002, MNRAS, 337, 1068

\bibitem{perlmutter} Perlmutter S., et al., 1999, ApJ, 517, 565

\bibitem[]{} Pointecouteau E., \& Silk J., 2005, MNRAS, in press, astro-ph/0505017

\bibitem{qin} Qin B., Wu, X.P., Zou Z.L., 1995, A\&A, 296, 264

\bibitem{ref4} Refregier R., 2003, ARA\&A, 41, 645.

\bibitem{reiss} Riess A., et al., 1998, AJ, 116, 1009

\bibitem{Romanowsky} Romanowsky A. J. et al. 2003, Science, 301, 1696

\bibitem{} Sanders R., 1999, ApJ, 512, L23

\bibitem{} Sanders R., 2003, MNRAS, 342, 901

\bibitem{} Sanders R., 2005, MNRAS, 363, 459

\bibitem{SandersMcGaugh} Sanders, R. , McGaugh S. 2002, ARA\&A, 40, 263,

\bibitem[]{} Shan H.Y., et al., 2005, in prep.

\bibitem[]{} Skordis C. et al., 2005, Phys. Rev. Lett. in press, astro-ph/0505519

\bibitem{spergel} Spergel D., et al., 2003, ApJSupp, 148, 175


\bibitem{Tegmark} Tegmark M., et al.,  2004, Phys. Rev. D, 69, 103501

\bibitem{vw2} Van Waerbeke L., Mellier Y., astro-ph/0305089.

\bibitem{} Worthey G., 1994, ApJS, 95, 107

\bibitem[]{} Zhao H., 2005, A\&A Letters, 444, L25
\bibitem[]{} Zhao H., \& Famaey B., 2005, ApJ, submitted (astro-ph/0512425)
\bibitem[]{} Zhao, H. S., Pronk, D., 2001, MNRAS, 320, 401
\bibitem[]{} Zhao H., Tian L., 2005, A\&A, submitted
\bibitem[]{} Zhao H., \& Qin B., 2005, ChJAA, accepted 

\end{thebibliography}
\end{document}